\documentclass[final,1p,times]{elsarticle}

\usepackage{amssymb}
\usepackage{amsmath}
\usepackage{caption}
\usepackage{subcaption}

\newtheorem{thm}{Theorem}

\newtheorem{defn}{Definition}[section]
\newdefinition{rmk}{Remark}
\newproof{pf}{Proof}
\newproof{pot}{Proof of Theorem \ref{thm2}}

\makeatletter

\def\ps@pprintTitle{%
 \let\@oddhead\@empty
 \let\@evenhead\@empty
 \def\@oddfoot{\footnotesize\itshape
 \ifx\@journal\@empty 
\else\@journal\fi\hfill\today}%
 \let\@evenfoot\@oddfoot}
\makeatother

\begin{document}

\begin{frontmatter}



\title{Vibration Control Design for Nonlinear Systems using Frequency Response Function}


\author{Suresh Thenozhi}
\author{Yu Tang}

\date{\today}
\address{Faculty of Engineering, National Autonomous University of Mexico, Mexico City, Mexico}

\begin{abstract}

A nonlinear frequency response based adaptive vibration controller is proposed for a class of nonlinear mechanical systems. In order to obtain the nonlinear Frequency Response Function (FRF), the convergence properties of the system are studied by using the convergence (contraction) theory. If the system under consideration is: 1) convergent, it directly enables to derive a nonlinear FRF for a band of excitation inputs, 2) non-convergent, first a controller is used to obtain the convergence and then the corresponding FRF for a band of excitation inputs is derived. Now the gains of the proposed adaptive controller are tuned such that a desired closed-loop frequency response, in the presence of excitation inputs is achieved. Finally, a building structure with nonlinear cubic stiffness and a satellite system are considered to illustrate the theoretical results.

\end{abstract}


\end{frontmatter}


\section{Introduction}
\label{}
Mechanical vibrations are present in countless real-life situations, where the mechanical systems exhibit oscillations when subjected to certain excitations. Most often these vibration phenomena are highly undesirable, which may even cause damage to the system itself. Vibration analysis and control are an active, vast, and growing research area, due to its practical importance and issues that arise in both linear and nonlinear system designs. The vibration control is concerned with the prediction and controlling of these undesired oscillations. The vibration can generally be controlled by adding controlling devices like dampers, isolators, and actuators to the system. These devices are added in such a way that the system's properties are modified to a desired one \cite{Thenozhi:2013a}. Most of the vibration control methods work based on time-domain techniques. These methods lack in describing how a closed-loop system respond to the input excitation at different frequencies and magnitudes. Since the vibration is characterized by its frequency (or frequencies), amplitude, and phase, it is important to study the frequency response of these systems.

Frequency domain techniques for linear systems have led to significant progress in analysis, modeling, and controller design \cite{Tang:1993a, Tang:1993b, Tang:1995}. In reality, mechanical systems posses many physical properties such as material property, geometric nonlinearity, damping dissipation, and even due to boundary conditions, which lead to nonlinear vibration problems \cite{Kerschen:2006}. For that reason, these nonlinear systems do not posses simple oscillations as defined in the case of linear systems. These nonlinearities can result in some complex phenomenon, like jumping, chaos, secondary resonance, bifurcation, etc. \cite{Nayfeh:2008}. In those instances, the classic linear frequency response analysis tools are insufficient to describe the behavior of the nonlinear system adequately. 

The methods such as describing function (DF) lack in accuracy due to its approximation scheme \cite{Khalil:2002}. Nonlinear FRF such as the Generalized Frequency Response Function (GFRF) \cite{George:1959} is limited to second order due to its multi-dimensional characteristics. An extension of the GFRF, termed as Output Frequency Response Function (OFRF) was proposed in \cite{Lang:1996, Lang:1997}, which represents the relation between the system parameters and frequency response using  finite Volterra series. However, this technique fails to detect some of the nonlinear phenomena such as the subharmonics, jumping, etc. \cite{Billings:2013}. In \cite{Pavlov:2007}, it has been shown that an FRF can be found for a class of nonlinear systems termed as convergent systems. The convergence property implies that the system trajectories tend towards the unique bounded solution. If the system under consideration is convergent, it signifies: 1) the system is stable \cite{Slotine:1998}, 2) an FRF can be obtained due to the existence of a unique steady-state solution \cite{Pavlov:2007}. Compared to the DF, GFRF, and OFRF, this FRF gives an exact frequency response via numerical or experimental approach. 

The main objective of this paper is to introduce the potential of convergence analysis and nonlinear FRF in vibration control problems. This paper considers a nonlinear FRF based vibration analysis and control of a class of nonlinear mechanical systems. In order to derive the FRF, first the convergence properties of the system have been established theoretically. In the case of non-convergent systems, a feedback control is designed to achieve convergence. In terms of control, the controller gains are adapted based on the FRF of the system derived within the excitation band of interest, which assures a satisfactory performance over that band. The control scheme developed here is applied to a building structure with cubic stiffness nonlinearity and to a satellite system, and the corresponding dynamic responses of both the controlled and uncontrolled cases are numerically evaluated. 

\section{FRF of Nonlinear Convergent Systems}

For a system to be satisfactory, it is necessary to analyze its stability. In general a system's stability is analyzed by examining, whether the equilibrium points so determined are stable. The convergence analysis, inspired by fluid mechanics, is the extension of the stability properties of asymptotically stable linear time-invariant systems. Unlike the Lyapunov stability theorem which defines the stability with respect to the equilibrium points, the convergence in a convergent system implies that the state trajectories with different initial conditions will converge to a unique bounded solution \cite{Slotine:1998}.

\begin{defn}
Let a dynamical nonlinear system be described by the differential equation
\begin{equation}
\dot{x}=f(x,t)  \label{nl}
\end{equation}%
with $x\in \mathbb{R} ^{n}$ is the state vector, $t \in \mathbb{R}_+,$ and $f:\mathbb{R} ^{n}\times \mathbb{R}_+ \rightarrow \mathbb{R} ^{n}$ is a smooth nonlinear function. For the above system, a region of convergence (or contraction region), $\mathcal{X}$ is defined where the system's Jacobian matrix, $J(x)=\partial f/\partial x$ is uniformly negative definite. 
\end{defn}

The convergence properties of the system (\ref {nl}) can be verified by performing a coordinate transformation on $J(x)$. The resulting generalized Jacobian is defined as
\begin{equation}
\mathcal{J}=\left( \dot{\Upsilon}+\Upsilon J(x) \right) \Upsilon ^{-1}  \label{gen}
\end{equation}
where $\Upsilon \left( x,t\right) $ is a uniformly invertible square matrix. If $\mathcal{J}$ is uniformly negative definite
\begin{equation}
\mathcal{J} \leq -\lambda _{\max }I,\text{\qquad }\forall x\in \mathcal{X}
\subset \mathbb{R}^{n},\text{ }\forall t\in \mathbb{R}
\end{equation}
where $\lambda _{\max }$ is the largest eigenvalue of $\mathcal{J}$, then the transformed system (\ref {gen}) is convergent, which implies that all the solutions of the original system (\ref{nl}) converge exponentially to a single trajectory, independently of the initial conditions. If $\mathcal{J}$ is negative semi-definite, then the system is semi-convergent under some mild conditions similar in Barbalat's lemma, this implies that the solutions converge each other asymptotically. The global convergence or semi-convergence is achieved when $\mathcal{X}=\mathbb{R}^{n}$.


FRF is the characteristics of a system that describes its response to an input as the function of frequency. Consider a nonlinear time-invariant system, which is forced by the input excitation $w\in \mathcal{W}$
\begin{equation}
\begin{array}{c}
\dot{x}=f\left( x,w\right) \\ 
y=g(x)%
\end{array}
\label{nly}
\end{equation}%
where $y$ is the system output. 
\begin{defn}
If the system (\ref{nly}) is convergent, then there exists an uniformly bounded steady-state (UBSS) solution, if for any $\rho >0$ there exists $\sigma >0$ such that for any input $w\in \mathcal{W}$ the following implication holds:
\begin{equation}
\left\vert w\right\vert \leq \rho \text{\qquad }\forall t\in 
\mathbb{R} \implies \left\vert \overline{x}_{w}\right\vert \leq \sigma \text{\qquad}\forall
t\in \mathbb{R} \label{ubss}
\end{equation}
where $\overline{x}_{w}$ is the steady-state solution, which depends on $w$. For the convergent systems, this bounded solution $\overline{x}_{w} $ is unique and $\lim_{t\rightarrow \infty }\left\Vert \overline{x}_{w}(t)-x_{w}(t)\right\Vert =0$ for any $x_{0}\in \mathcal{X}$ and hence the system (\ref{nly}) is exponentially stable \cite{Pavlov:2007}.
\end{defn}

\begin{defn}
If the system (\ref{nly}) is convergent with UBSS property for a certain class of harmonic inputs $w(t)=a\sin (\omega t)\in \mathcal{W}$, then there exists a nonlinear function $\alpha : \mathbb{R}^{3}\rightarrow \mathbb{R}^{n}$ such that
\begin{equation}
\overline{x}_{w}(t):=\alpha \left( v_{1},v_{2},\omega \right) 
\end{equation}%
where $v_{1}=a\sin (\omega t)$ and $v_{2}=a\cos (\omega t)$. The nonlinear function $\alpha \left( v_{1},v_{2},\omega \right) $ is known as the state frequency response function and the function $g\left( \alpha \left( v_{1},v_{2},\omega \right) \right) $ is known as the output frequency response function, which relates the harmonic input to the corresponding steady-state output \cite{Pavlov:2007}. 
\end{defn}

The output of the forced system at steady-state can be expressed as $\overline{y}_{w}(t)=g(\overline{x} _{w}(t))$, which will have the same period of the input signal $w$, but not necessarily sinusoidal. Now the output response of the system for various amplitude and frequency inputs can be represented using an amplification gain $\gamma _{a}\left( \omega \right) $, which is the ratio between the maximal absolute value at steady-state and the corresponding input signal amplitude, so that 
\begin{equation}
\gamma _{a,\omega }=\frac{1}{a}\left(
\sup_{v_{1}^{2}+v_{2}^{2}=a^{2}}\left\vert g\left( \alpha \left(
v_{1},v_{2},\omega \right) \right) \right\vert \right) =\frac{\left\vert 
\overline{y}_{w}(t)\right\vert }{a}  \label{amp}
\end{equation}

Analyzing the above function, one can find the critical input amplitudes and frequencies for the system and an appropriate controller can be designed to bypass any undesirable effects such as the vibration. 

\section{Vibration Control of Nonlinear Mechanical Systems}
There are three basic elements in a mechanical system: mass which stores kinetic energy, spring which stores the potential energy and the damper which dissipates the energy. When an external force is applied to a structure, it produces change in its displacement, velocity, and acceleration.

Let us consider a nonlinear multi-degree-of-freedom mechanical system of the following form
\begin{equation}
M\ddot{q}+C\dot{q}+Kq+\Phi (q, \dot{q})=\Lambda w  \label{sys1}
\end{equation}
where $M,C,$ and $K\in \mathbb{R}^{n_q\times n_q}$ \footnote{In the case of second-order mechanical systems $n=2n_q$.} are the positive definite matrices correspond to the mass, damping, and stiffness respectively, $\ddot{q},\dot{q},$ and $q\in \mathbb{R}^{n_q}$ are the relative acceleration, velocity, and displacement vectors respectively, $\Lambda \in \mathbb{R} ^{n_q} $ denotes the influence of the input excitation force $w$ on the system, and $\Phi (q, \dot{q})\in \mathbb{R} ^{n_q}$ is the system nonlinearity vector. 

\subsection{Problem Formulation}

In vibration control applications, the system dynamics were modified favorably by adding passive or active devices. Most of the passive devices can be tuned only to a particular structural frequency and damping characteristics. But the nonlinearities in the systems cause variations in its natural frequencies and mode shapes. Since the passive devices cannot adapt to these structural response changes, it cannot assure a successful vibration suppression. These shortcomings can be overcome by using an active control system, where the system's output response is measured using sensors and an appropriate control force, calculated by a pre-assigned controller is used to drive the actuators for suppressing the unwanted structural vibration. The mechanical system with an active control system can be represented as,
\begin{equation}
M\ddot{q}+C\dot{q}+Kq+\Phi (q,\dot{q})=\Lambda w+\Gamma u  \label{uu}
\end{equation}%
where $u\in \mathbb{R}^{n_{u}}$ is the control signal generated by a control algorithm by means of which the system can be controlled and $\Gamma \in \mathbb{R} ^{n_q \times n_u} $ is the actuator location matrix.

Now let us consider a feedback controller of form
\begin{equation}
u=-\Theta x \label{cont}
\end{equation}
where $\Theta \in \mathbb{R} ^{n_u \times n}$ is the state feedback controller gain matrix and $x=\left[q^T \text{ } \dot{q}^T\right]^T$. The control objective is to find a frequency response based adaptive control rule for the controller gain
\begin{equation}
\dot{\Theta}=\Psi \left( \mathcal{F}\right)   \label{adap1}
\end{equation}
where $\mathcal{F}$ is the FRF of the system, such that a desired vibration attenuation by means of a specified frequency response is achieved for the closed-loop system (\ref{uu}).

\subsection{Vibration Control of Convergent Systems}

In this section, first we will establish the convergence property of a nonlinear mechanical system (\ref{sys1}), which will help us to derive the FRF of the open-loop system. Next we will derive the range of controller gains for which the closed-loop system is convergent, hence $\Theta \in \Omega_{\Theta} \subset \mathbb{R}^{\Theta}$, where $\Omega_{\Theta}$ is a compact set. Finally, based on the system's FRF, an adaptive algorithm of form (\ref{adap1}) will be presented to tune the controller gains $\Theta $ within a convergence region such that the vibration is minimized.

\subsubsection{Convergence Analysis}
Let us consider the multi-degree-of-freedom mechanical system shown in (\ref{sys1}), with an odd polynomial nonlinearity of form
\begin{equation}
\Phi (q)= \left[\phi _{1}(q_1),...,\phi _{n_q}(q_{n_q})\right]^T \in \mathbb{R} ^{n_q}
\label{odd1}
\end{equation}%
where
\begin{equation*}
\phi _{i} (q_i)=\sum \limits_{p=1}^{\overline{p}}b_{i,2p+1}q^{2p+1} _i,\text{\qquad }%
b_{i,2p+1}>0, \text{ } i=1,...,n_q \label{odd2}
\end{equation*}
with $\overline{p}$ as the highest order of the nonlinearity. By letting $q=x_{1}\in \mathbb{R}^{n_q}$ and $\dot{q}=x_{2}\in \mathbb{R}^{n_q}$, the corresponding second order model represented can be written as a set of first order differential equations
\begin{equation}
\left[ 
\begin{array}{c}
\dot{x}_{1} \\ 
\dot{x}_{2}%
\end{array}%
\right] =\left[ 
\begin{array}{c}
x_{2} \\ 
-M^{-1}\left[ Cx_{2}+Kx_{1}+\Phi (x_{1})-\Lambda w\right]%
\end{array}%
\right]  \label{ss}
\end{equation}%
which is equivalent to the nonlinear model (\ref{nly}), where the nonlinear vector functions $f(x,w)$ are
\begin{equation}
\left[ 
\begin{array}{c}
f_{1}(x_{1},x_{2},w) \\ 
f_{2}(x_{1},x_{2},w)%
\end{array}%
\right] =\left[ 
\begin{array}{c}
x_{2} \\ 
-M^{-1}\left[ Cx_{2}+Kx_{1}+\Phi (x_{1})-\Lambda w\right]%
\end{array}%
\right]
\end{equation}

\begin{thm}
The class of mechanical system with odd polynomial nonlinearity, represented in (\ref{sys1}) is convergent.
\begin{pf}
The system (\ref{nl}) can be represented in terms of virtual displacement $%
\delta x$\ as \cite{Slotine:1998}
\begin{equation}
\delta \dot{x}=J(x)\delta x  \label{vir}
\end{equation}%
The Jacobian matrix of the system represented in (\ref{ss}) is
\begin{equation}
J(x)=\left[ 
\begin{array}{cc}
0_{n_q \times n_q} & I_{n_q \times n_q} \\ 
-M^{-1}\left[ K+\Phi _{J}(x_{1})\right]  & -M^{-1}C%
\end{array}%
\right] \in \mathbb{R} ^{n \times n}   \label{J}
\end{equation}%
where $\Phi _{J}(x_{1})=\Phi^{T} _{J}(x_{1})=\frac{\partial \Phi (x_{1})}{\partial x_1}\geq 0$. Generally, it is difficult to obtain a uniformly negative definite Jacobian matrix directly for the system represented in the form (\ref{J}). However, a transformation can always be done to prove the negative definiteness of the Jacobian matrix. Consider the following coordinate transformation
\begin{equation}
\delta z=\Upsilon \delta x
\end{equation}%
where the transformation matrix$\ \Upsilon $ is%
\begin{equation}
\Upsilon =\left[ 
\begin{array}{cc}
I_{n_q \times n_q} & 0_{n_q \times n_q} \\ 
I_{n_q \times n_q} & I_{n_q \times n_q}
\end{array}%
\right]  \label{T}
\end{equation}%
then 
\begin{equation}
\left[ 
\begin{array}{c}
\delta z_{1} \\ 
\delta z_{2}%
\end{array}%
\right] =\left[ 
\begin{array}{c}
\delta x_{1} \\ 
\delta x_{1}+\delta x_{2}%
\end{array}%
\right]
\end{equation}

Using (\ref{gen}) the generalized Jacobian $\mathcal{J}$ can be calculated as
\begin{eqnarray}
\mathcal{J} &=&\Upsilon J\left( x\right) \Upsilon ^{-1}  \notag \\
&=&\left[ 
\begin{array}{cc}
-I_{n_q \times n_q} & I_{n_q \times n_q} \\ 
-I_{n_q \times n_q}-M^{-1}\left[ K+\Phi _{J}(z_{1})-C\right] & I_{n_q \times n_q}-M^{-1}C%
\end{array}%
\right]
\end{eqnarray}

Now the transformed system is 
\begin{equation}
\delta \dot{z}=\mathcal{J}\delta z  \label{vir1}
\end{equation}

The squared distance between the trajectories of the virtual system can be calculated as
\begin{eqnarray}
\frac{d}{dt}\left\Vert \delta z\right\Vert &=&2\delta z^{T}\delta \dot{z} 
\notag \\
&=&2\delta z^{T}\mathcal{J}\delta z  \notag \\
&\leq &\lambda _{\max }\left( \mathcal{J}\right) \left\Vert \delta
z\right\Vert  \label{con}
\end{eqnarray}%

In order to achieve exponential convergence, $\mathcal{J}$ must be negative definite, which is established since $\mathcal{J}_{11}<0$ and $K+\Phi _{J}(z_{1})>0$ $(K>0$ and $\Phi_{J}(z_{1})\geq 0)$. Hence, the solutions of the virtual system (\ref{vir1}) converge exponentially to zero with a rate $\lambda _{\max }\left( \mathcal{J}\right)$, which implies that the solutions of the actual system (\ref{sys1}) also converge to each other. 

\end{pf}
\end{thm}

\subsubsection{PD Controller Design}

PD control may be the simplest controller for the vibration control system, which provides high robustness with respect to uncertainties. It has the following form
\begin{equation}
u=-\Theta _{p}\ e-\Theta _{d}\dot{e}  \label{pd1}
\end{equation}%
where $\Theta _{p},\Theta _{d} \in \mathbb{R}^{n_q \times n_q}$ \footnote{In the case of full-state feedback control $n_q=n_u$. In that case $\Gamma$ is an Identity matrix of dimension $n_q$, hence ignored in the later part of this paper.} are symmetric matrices correspond to the proportional and derivative gains, respectively, $e=\left(q-q^{d}\right)$ corresponds to the position error, $\dot{e}=\left(\dot{q}-\dot{q}^{d}\right)$ corresponds to the velocity error, and $q^{d}$ is the desired position. In active vibration control case, the references are $q^{d}=\dot{q}^{d}=0$, hence (\ref{pd1}) becomes
\begin{equation}
u=-\Theta _{p}q-\Theta _{d}\dot{q}  \label{pd}
\end{equation}

The closed-loop system (\ref{uu}) with the PD controller (\ref{pd}) is 
\begin{equation}
M\ddot{q}+C\dot{q}+Kq+\Phi (q)=\Lambda w - \Theta _{p}q - \Theta _{d}\dot{q}  \label{cl}
\end{equation}%
which can be written in the state-space form as
\begin{equation}
\left[ 
\begin{array}{c}
\dot{x}_{1} \\ 
\dot{x}_{2}%
\end{array}%
\right] =\left[ 
\begin{array}{c}
x_{2} \\ 
-M^{-1}\left[ 
\begin{array}{c}
\left( C+ \Theta _{d}\right) x_{2}+ \left( K+ \Theta _{p}\right) x_{1}+\Phi (x_{1})- \Lambda w
\end{array}
\right] 
\end{array}
\right] \label{cl2}
\end{equation}

The range of gains for which the system is convergent is defined in the following theorem.

\begin{thm}
Consider the class of nonlinear system (\ref{uu}) with an external excitation $w$ controlled using the control law (\ref{cont}). If we choose the control gains such that $\Theta=\left[\Theta_p \text{ } \Theta_d \right] \in \Omega_{\Theta} \subset \mathbb{R}_{+}^{\Theta}$, then the state trajectories of the closed-loop system (\ref{uu}) lies within the region of convergence $\overline{x}_w \in \mathcal{X}$ and hence exponentially stable.
\begin{pf}
The Jacobian matrix of the closed-loop system (\ref{cl2}) is%
\begin{equation}
J(x)=\left[ 
\begin{array}{cc}
0_{n_q \times n_q} & I_{n_q \times n_q} \\ 
-M^{-1}K_{K} & -M^{-1}K_{C}%
\end{array}%
\right]   \label{Jc}
\end{equation}%
where $K_{K}=K+ \Theta _{p}+\Phi _{J}(x_{1})$ and $K_{C}=C+ \Theta _{d}$. The generalized Jacobian $\mathcal{J}$ can be now found as 
\begin{eqnarray}
\mathcal{J} &=&\Upsilon J\left( x\right) \Upsilon ^{-1}  \notag \\
&=&\left[ 
\begin{array}{cc}
-I_{n_q \times n_q} & I_{n_q \times n_q} \\ 
-I_{n_q \times n_q}-M^{-1}\left[ K_{K}-K_{C}\right]  & I_{n_q \times n_q}-M^{-1}K_{C}%
\end{array}%
\right] 
\end{eqnarray}
Since $K_{K}>0$, the matrix $\mathcal{J}$ is negative definite. From the above discussions and from (\ref{ubss}), it can be concluded that for a bounded input excitation $\left\vert w\right\vert \leq \rho $, the controller with positive proportional and derivative gains will result an exponentially stable solution, hence $\overline{x}_w \in \mathcal{X}$.
\end{pf}
\end{thm}

\subsubsection {Nonlinear Frequency Response based Adaptive Controller Design} \label{frfa}

Adaptive techniques are widely used for active vibration control applications. Traditional time-domain based adaptive schemes update the controller gains based on the error at that moment. However, they cannot assure a satisfactory vibration attenuation for a given band of excitation. Here, designing of an adaptive control algorithm based on the system's FRF is discussed. 

The amplification gain $\gamma _{a}\left( \omega \right)$ of the system, for a range of magnitudes ($a\in \left( \underline{a},\overline{a} \right) \in \mathbb{R}^{r}$) and frequencies ($\omega \in \left( \underline{\omega },\overline{\omega }\right) \in \mathbb{R}^{s}$) can be represented as
\begin{equation}
\mathcal{F}_{0}=\left[ 
\begin{array}{ccc}
\gamma _{\underline{a},\underline{\omega }} & \ldots  & \gamma _{\underline{a%
},\overline{\omega }} \\ 
\vdots  & \ddots  & \vdots  \\ 
\gamma _{\overline{a},\underline{\omega }\text{ }} & \ldots  & \gamma _{%
\overline{a},\overline{\omega }\text{ }}%
\end{array}%
\right] \in 
\mathbb{R}
^{r\times s}  \label{frf}
\end{equation}

The above matrix can be considered as the open-loop FRF matrix of the system. Now $\mathcal{F}_{0}$ can be analyzed in order to get a knowledge about the critical magnitudes and frequencies of the excitation input, at which the system possess a larger amplification gain. One way to evaluate the FRF matrix is by finding its Frobeinus norm (F-norm), which are sensitive towards its each elements. F-norm of a matrix $A\in \mathbb{R}^{\overline{m}\times \overline{n}}$ can be calculated as
\begin{equation}
\left\Vert A\right\Vert _{F}=\sqrt{\sum\limits_{i=1}^{\overline{m}%
}\sum\limits_{j=1}^{\overline{n}}\left\vert a_{ij}\right\vert ^{2}}=\sqrt{\textup{%
tr}\left( A^{T}A\right) }\geq 0  \label{norm}
\end{equation}

For the critical input excitations, the F-norm of the open-loop system FRF $\left\Vert \mathcal{F}_{0}\right\Vert _{F}$ will be higher. A PD controller can be used to reduce the system peaks at those critical points. The proportional and derivative gains provide virtual stiffness and damping to the closed-loop system, respectively. The damping term helps in reducing the resonance peaks by dissipating the vibration energy. Once we increase $\Theta_d$, the $\lambda_{max}$ of the closed-loop system becomes more negative and as a result the convergence rate ($e^{\lambda_{max}}$) increases accordingly. On the other hand $\Theta_p$ can be tuned such that the system resonance frequency can be shifted beyond the frequency band of potential excitation. The stiffness value must be chosen carefully without causing any undesirable effects at other excitation inputs. This phenomenon is termed as waterbed effect, where the reduction of the magnitude response of the closed-loop system in one frequency range will result in the increase of the magnitude response in some other frequency range \cite{Doyle:1992}. By analyzing the system's closed-loop FRF, the controller gains can be tuned in order to achieve maximal vibration attenuation at the critical excitations without causing undesirable effects at other operational frequencies.

The steady-state output of the closed-loop system for a particular value of proportional and derivative gains can be represented as $\overline{y}_{w}(t,\Theta)$. The amplification gain of the closed-loop system, denoted by $\gamma _{a,\omega}\left( \Theta \right)$, satisfies the following relation
\begin{equation}
\left\vert \overline{y}_{w}(t,\Theta )\right\vert =\gamma _{a,\omega }\left(
\Theta \right) \left\vert a\right\vert  \label{ampt}
\end{equation}

Hence the peak vibration output of the system can be attenuated by minimizing the amplification gains. Using the amplification gains obtained under a range of excitation, the closed-loop FRF matrix, $\mathcal{F}_{\Theta }$ can be constructed similar to (\ref{frf}). The control objective is to minimize the FRF magnitude such that
\begin{equation}
\left\Vert \mathcal{F}_{\Theta }\right\Vert _{F}=\left(
\sum\limits_{i=1}^{r}\sum\limits_{j=1}^{s}\left\vert \gamma _{a,\omega
}\left( \Theta \right) _{i,j}\right\vert ^{2}\right) ^{1/2}\leq \delta
<\left\Vert \mathcal{F}_{0}\right\Vert _{F}  \label{min}
\end{equation}%
where $\delta $ is the acceptable vibration range. For the ideal vibration attenuation case $\delta \ $may be chosen as zero, which means $\gamma _{a,\omega }\left( \Theta \right) =0$, due to the property $\left\Vert \mathcal{F}_{\Theta }\right\Vert _{F}=0 \implies \mathcal{F}_{\Theta }=0$. However, in practice it is not possible to remove the entire vibration from the system due to the waterbed effect and limitations of the control devices. Our goal is to choose a $\delta $ according to the practical situations and to achieve it by using an adaptive scheme.

Here, the design requirement is that each increment in the controller gains at least minimizes the F-norm of the FRF matrix. The controller gains are adapted by using the following algorithm
\begin{equation}
\Delta \Theta _{i} \triangleq \Theta _{i+1}-\Theta _{i}=\Gamma_{\Theta} \left\Vert 
\mathcal{F}_{\Theta ,i}\right\Vert _{F}\varepsilon _{i}  \label{adap}
\end{equation}%
where
\begin{equation*}
\Gamma_{\Theta} =\Gamma ^{T} _{\Theta}=\left[ 
\begin{array}{cc}
\Gamma _{\Theta _{p}} & 0 \\ 
0 & \Gamma _{\Theta _{d}}%
\end{array}%
\right] >0
\end{equation*}
is the adaptation step size,
\begin{equation*}
\left\Vert \mathcal{F}_{\Theta ,i}\right\Vert _{F}=\left[ 
\begin{array}{cc}
\left\Vert \mathcal{F}_{\Theta ,i}\left( \overline{x}_{1}\right) \right\Vert _{F} & 0 \\ 
0 & \left\Vert \mathcal{F}_{\Theta ,i}\left( \overline{x}_{2}\right) \right\Vert _{F}%
\end{array}%
\right] ,
\end{equation*}
and $\varepsilon_{i} =\left[ \left\Vert \mathcal{F}_{\Theta ,i}\left( \overline{x}_{1}\right)
\right\Vert _{F}-\delta _{\overline{x}_{1}},\left\Vert \mathcal{F}_{\Theta ,i}\left(
\overline{x}_{2}\right) \right\Vert _{F}-\delta _{\overline{x}_{2}}\right] ^{T}$ is the error between the F-norm of the closed-loop FRF at $i$-th iteration and the desired range $\delta$. If the gains are chosen such that
\begin{equation}
\Theta _{i}=\left\{ 
\begin{array}{ccc}
\Theta _{i} & \textup{if} & \Theta _{i}>\Theta _{\min } \\ 
\Theta _{\min } & \textup{if} & \Theta _{i}\leq \Theta _{\min }%
\end{array}%
\right.  \label{proj}
\end{equation}%
where $\Theta _{\min}=\left[\Theta _{p \min} \text{ } \Theta _{d \min}\right]$ is used to assure that the controller gain $\Theta _{i}$ is always positive, then the solutions of the closed-loop system stays within the convergence region.

Equation (\ref{adap}) is used to calculate the new controller gains, according to its previous FRF matrix. When the error $\varepsilon $ is positive, the gains will increase and for negative $\varepsilon $ the gain decreases. Based on the $\varepsilon $, the gains will adapt over each iteration until a satisfactory vibration attenuation is achieved for a band of excitation, that is $\left\Vert \mathcal{F}_{\Theta }\right\Vert _{F}=\delta $, hence $\varepsilon _{i}\rightarrow 0$ as $i\rightarrow \infty $. Moreover, the projection operator (\ref{proj}) will assure that the closed-loop system (\ref{cl}) with the adaptation scheme (\ref{adap}) operates within the convergence region. 

\subsection{Vibration Control of Non-Convergent Systems}
\label{snc}
In the previous case the open-loop configuration of the system under consideration is already convergent. But in practice, there exist mechanical systems which are not exponentially convergent. 
Let us consider a nonlinear mechanical system of form
\begin{equation}
{H}(q) \ddot{q}+ {C}(q, \dot{q}) \dot{q} = \tau
\label{nc}
\end{equation}
where $H(q) \in \mathbb{R}^{n_q\times n_q}$ is the symmetric positive definite inertia matrix, $C(q, \dot{q}) \dot{q} \in \mathbb{R}^{n_q}$ is a nonlinear vector function of $q$ and $\dot{q}$ due to the Coriolis and centripetal effects, and $\tau \in \mathbb{R}^{n_q}$ is the actuator input torque. Since, for the above system the gravitational force term $\left(\partial U/\partial q\right)^T$ corresponds to the potential energy $U$ is absent, and the corresponding total energy of the system $E$, is equal to its kinetic energy, which is
\begin{equation}
E = \frac{1}{2} \dot{q}^T H \dot{q}
\label{E}
\end{equation}
and the corresponding variation of the energy is 
\begin{eqnarray}
\dot{E}&=&\dot{q}^T H \ddot{q}+\frac{1}{2} \dot{q}^T \dot{H} \dot{q} \notag \\
&=&\dot{q}^T \left(-{C} \dot{q}+\tau \right)+\frac{1}{2} \dot{q}^T \dot{H} \dot{q} \notag \\
&=&\frac{1}{2}\dot{q}^T \left(\dot{H}-2{C} \dot{q}\right)+ \dot{q}^T \tau 
\end{eqnarray}
Considering $\tau=0$ and by using the skew-symmetric property of the system (\ref{nc}) such that
\begin{equation}
 q^T\left[\dot {H}(q) - 2 C(q, \dot{q}) \right]q=0
\label{skew}
\end{equation}
yields $\dot{E}=0$, which implies the total system's energy is constant due to the conservation of energy. This is due to the fact that the matrix $C(q, \dot{q})$ does not contain any dissipative forces. Now the non-conservative force, like the input torque ($\dot{E}=\dot{q}^T \tau$), might be designed such that system trajectories converges exponentially. In this section, we discuss a method to provide exponential convergence by means of a feedback control loop. 

\subsubsection{Obtaining Convergence}

One way to achieve the system convergence is by introducing a virtual stiffness and damping term by means of a standard PD controller. Applying the PD control law in (\ref{nc}) yields 
\begin{equation}
{H}(q) \ddot{q}+ {C}(q, \dot{q}) \dot{q} + K_D \dot{q}= -K_P q
\label{tau1}
\end{equation}

Now let us define a local coordinate system $z$ defined as 
\begin{equation}
\dot z = H \delta v + C \delta q + K_D \delta q = -K_P q
\label{zz}
\end{equation}

The virtual displacement of the above system can be written as 
\begin{equation}
\delta z = H \delta v + C \delta q + K_D \delta q
\label{}
\end{equation}
and the corresponding dynamic variation is 
\begin{equation}
\delta \dot{z} = -K_P \delta q
\label{}
\end{equation}

Now the convergence can be established as follows,
\begin{eqnarray}
\frac{d}{dt}\left[ \delta v^{T}K_{P}^{-1}\delta v+\delta q^{T}H\delta q%
\right]  &=&-2\delta v^{T}\delta q+2\delta q^{T}\left( H\delta v+C\delta
q\right)  \notag \\
&=&-2\delta q^{T}K_{D}\delta q\leq 0
\end{eqnarray}

Using Barbalat's lemma it can be shown that both the $\delta v$ and $\delta q$ will converge to each other asymptotically, which implies the semi-convergence of the original system (\ref{nc}). For that reason the system FRF cannot be obtained, which requires that the system states reach steady-state exponentially. 

Now let us consider an energy-based controller of form \cite{Slotine:2004}
\begin{equation}
\tau = {H}(q) \ddot{q}^r+ C(q, \dot{q}) \dot{q}^r - K_{r}(\dot{q}-\dot{q}^r)
\label{ue}
\end{equation}
where $K_{r}$ is a symmetric positive definite gain matrix and the reference velocity
\begin{equation*}
\dot{q}^r=\dot{q}^d-\Lambda _r e, \text{\qquad } \Lambda _r>0
\end{equation*}
and the corresponding tracking error $r$ is
\begin{equation}
r=\dot{e}+\Lambda _r e= \dot{q}-\dot{q}^r
\end{equation}

The virtual system is 
\begin{equation}
\tau = {H}(q) \dot{z}+ {C}(q, \dot{q}) z - K_{r}(\dot{q}-z)
\label{vir2}
\end{equation}

To examine the convergence properties, the virtual distance can be calculated as
\begin{eqnarray}
\frac{d}{dt}\left[\delta z^{T}H\delta z\right]  &=&-2\delta z^{T}\left(H \ddot{q}-H \ddot{q}^{r}\right)+\delta z^{T} \dot{H} \delta z \notag \\
&=&-2\delta z^{T}\left(-{C}\dot{q}+\tau-H \ddot{q}^{r}\right)+\delta z^{T} \dot{H} \delta z \notag \\
&=&-2\delta z^{T}\left(-H \ddot{q}^{r}-{C} \dot{q}^r+\tau \right)+\delta z^{T} \left(\dot{H} - 2{C} \right) \delta z 
\end{eqnarray}
 
Once again using the skew-symmetric property (\ref{skew}) of the system (\ref{nc}) and the controller (\ref{ue}), yields
\begin{equation}
\frac{d}{dt}\left[\delta z^{T}H\delta z\right]=-2\delta z^{T}K_{r}\delta z<0
\label{}
\end{equation}
which indicates that $\dot{q}$ converges to $\dot{q}^r$ exponentially. By noting that the trajectories of (\ref{nc}) and (\ref{ue}) are particular solutions of the virtual system (\ref{vir2}) its convergence signifies that $q$ tends to $q_d$ exponentially. Now an active vibration controller can be used to minimize the vibration caused by excitation inputs.

\subsubsection{PD Controller Design}

In order to analyze the convergence property the PD controller law can be rewritten as 
\begin{equation}
u=-\Theta_r r=-\Theta_r\left(\dot{e}+\Lambda _r e \right)
\label{pdr}
\end{equation}
where $\Theta_p=\Theta_r \Lambda_r, \text{ } \Theta_d=\Theta_r$.
Now the energy-based controller (\ref{ue}) along with PD controller (\ref{pdr}) can be represented as 
\begin{equation}
\tau = {H}(q) \ddot{q}^r+ {C}(q, \dot{q}) \dot{q}^r - K_{r}(\dot{q}-\dot{q}^r)- \Theta_r(\dot{q}-\dot{q}^r)
\label{ute}
\end{equation}

\begin{thm}
The nonlinear mechanical system (\ref{nc}) with energy-based controller (\ref{ue}) is convergent for the PD controller (\ref{pdr}), such that  $\Theta \in \Omega_{\Theta} \subset \mathbb{R}_{+}^{\Theta}$.
\begin{pf}
The virtual system is 
\begin{equation}
\tau = {H}(q) \dot{z}+ {C}(q, \dot{q}) z - K_{r}(\dot{q}-z)-\Theta_r(\dot{q}-z)
\end{equation}

Using the controller (\ref{ute}) the virtual distance can be calculated as
\begin{equation}
\frac{d}{dt}\left[\delta z^{T}H\delta z\right]=-2\delta z^{T}\left(K_{r}+\Theta_r\right)\delta z
\label{}
\end{equation}
which signifies that $\dot{q}$ converges to $\dot{q}^r$ exponentially, therefore, exponential convergence of $q$ to $q_d$ is guaranteed. 
\end{pf}
\end{thm}

Since the controller in (\ref{pdr}) has the structure and gain ranges similar to that of (\ref{pd}), we are in a position to implement the frequency response based adaptive controller proposed in Section \ref{frfa} to the system (\ref{nc}) with controller (\ref{ue}).

\section{Results and Discussions}

The frequency response based vibration control algorithm proposed in this paper can be used in a variety of vibration control applications. In this section, we have considered two important vibration control applications: 1) a building structure with cubic stiffness, which is a convergent system, 2) a satellite system, which is a non-convergent system. The performance of the proposed algorithm was evaluated via numerical simulations. 

\subsection{Application to Building Structures}

Protection of large civil structures and human occupants from natural hazards like an earthquake and wind is very important vibration control application. Many attempts have been made to introduce advanced controllers for the active vibration control of building structures \cite{Thenozhi:2013a}. Most of these methods work based on time-domain techniques. In practice, the building structures behave nonlinearly under large deformations, which can happen during strong seismic events. During these situations the well established linear frequency response analysis tools cannot be applied.

\begin{figure}[h]
\begin{center}
\includegraphics[width=7cm]{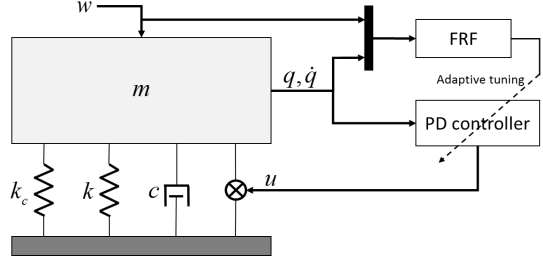}    
\caption{Block diagram of the proposed active vibration control system.} 
\label{block}
\end{center}
\end{figure}

The cubic nonlinearity have been chosen here, which have received considerable interest in literature. In order to present the main idea, let us consider a single-degree-of-freedom mechanical system with a cubic stiffness element ($k_{c}$)
\begin{equation}
m\ddot{q}+c\dot{q}+kq+k_{c}q^{3}=w  \label{cub}
\end{equation}
where the parameters are set as: $m=1 \textup{kg},c=0.4\textup{N} \textup{s}/\textup{m}$, and $k=k_{c}=36\textup{N}/\textup{m}$. The control programs were operated in Windows 7 with Matlab 8.0/Simulink. All the control actions were employed at a sampling period of $10\textup{ms}$. The magnitudes and frequencies of the input excitation are $a(\textup{N})\in (0.5,6)$ and $\omega (\textup{rad}/\textup{s})\in (3,9)$. The block diagram of the implementation of the proposed algorithm is shown in Figure \ref{block}. 

\begin{figure}[h]
\begin{center}
\includegraphics[width=7cm]{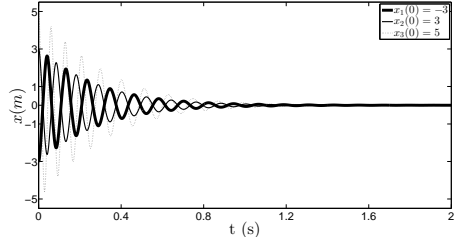}    
\caption{Convergence of trajectories for different initial conditions.} 
\label{init}
\end{center}
\end{figure}

The convergence of solutions for different initial conditions are shown in Figure \ref{init}. From this plot, it can be seen that the solutions for different initial conditions $\left(x_1(0)=-3, x_2(0)=3, x_3(0)=5\right)$ converge to the equilibrium point. Due to this convergence property, the FRF from the steady-state response can be derived without considering different initial conditions.

\begin{figure}
       \centering
        \begin{subfigure}[h] {0.35\textwidth}
                \includegraphics[width=\textwidth]{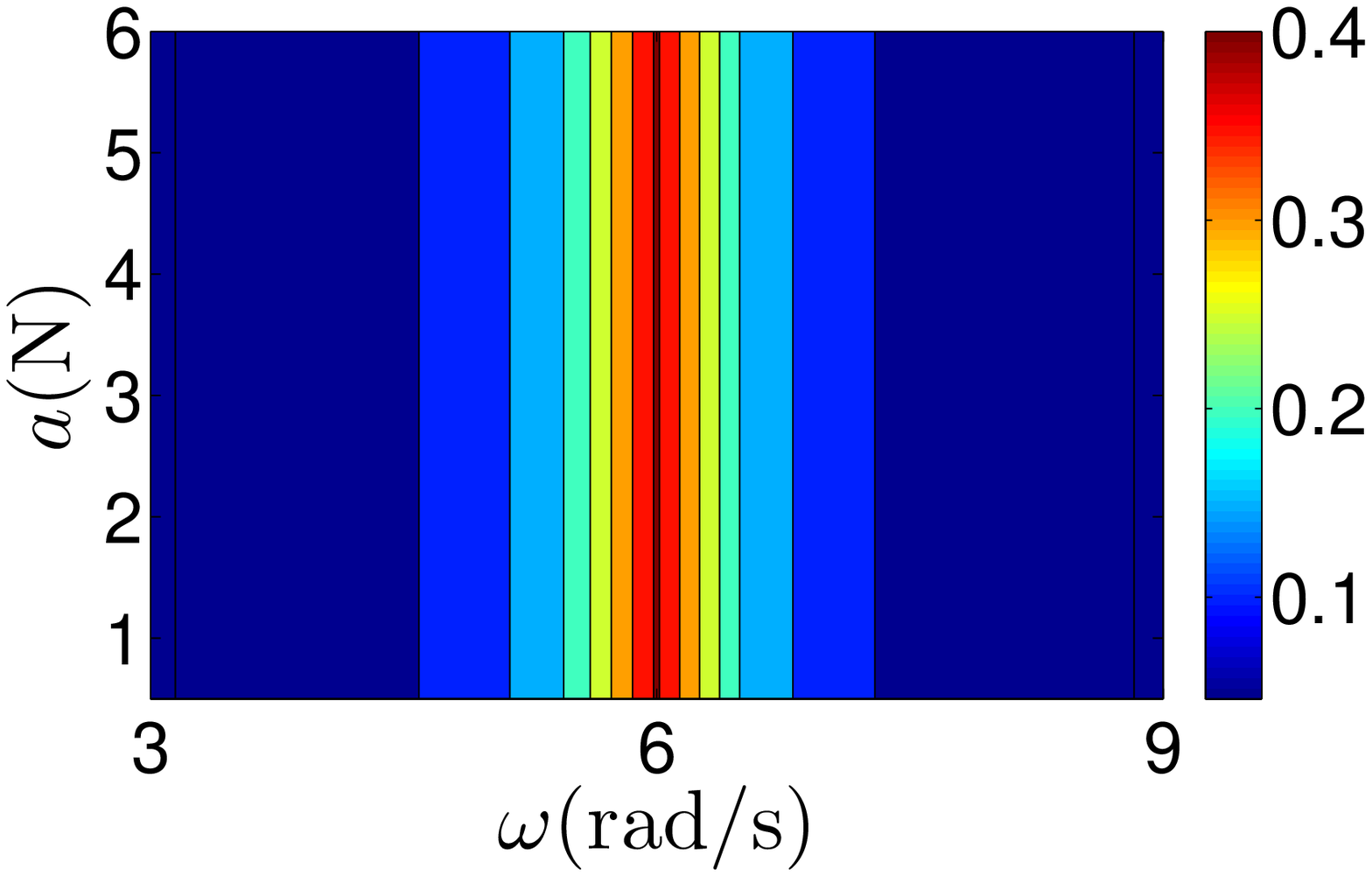}
                \caption{Linear case $(k_{c}=0)$}
                \label{flin1}
        \end{subfigure}%
        \begin{subfigure}[h] {0.35\textwidth}
                \includegraphics[width=\textwidth]{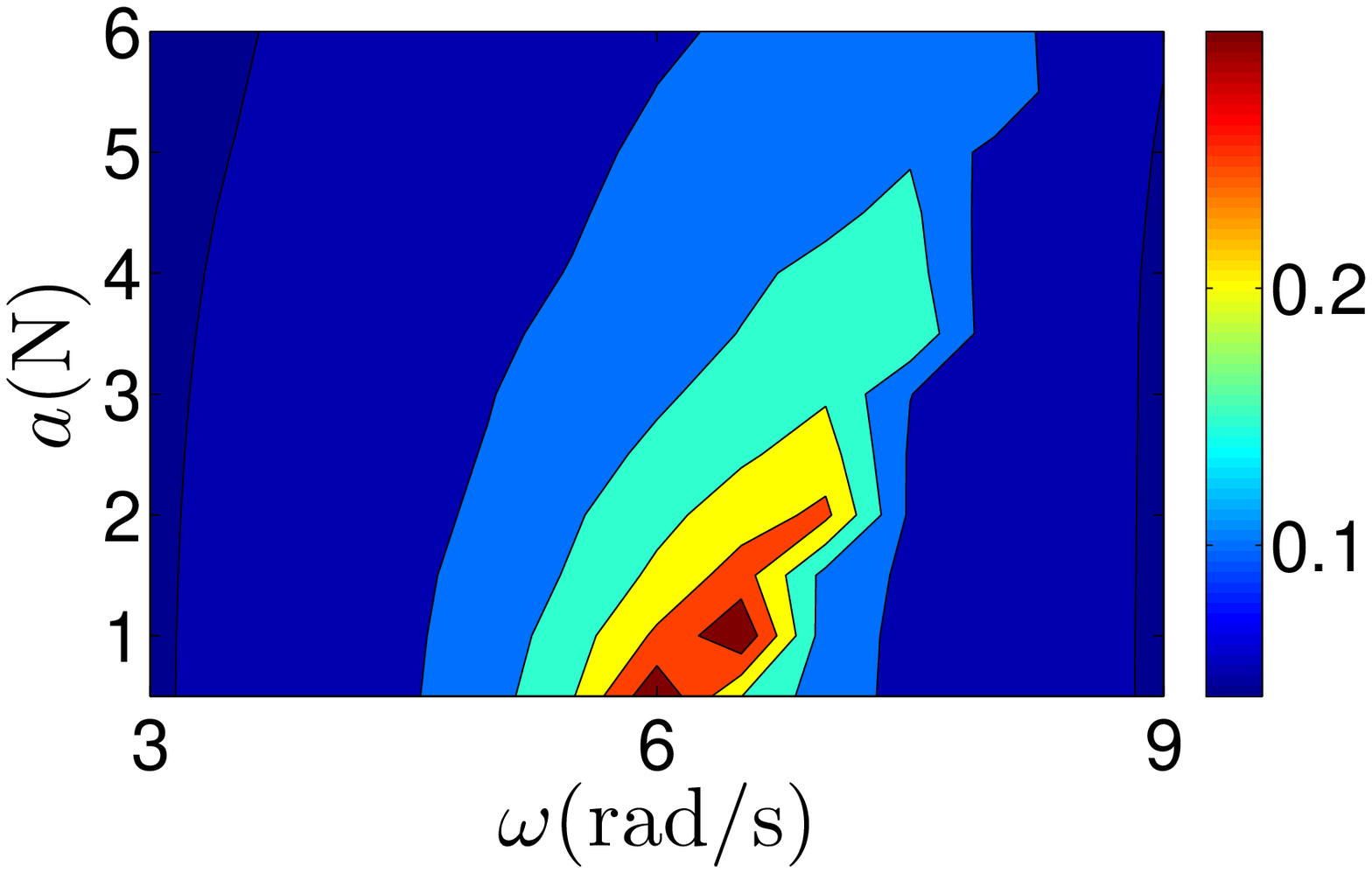}
                \caption{Nonlinear case $(k_{c}=100)$}
                \label{fnlin1}
        \end{subfigure}
       \caption{Open-loop FRF $\left(\mathcal{F}_{0}\right)$ of the mechanical system.}\label{F0}
\end{figure}

Figure \ref{F0} shows counter plots of the open-loop FRF of the mechanical system without and with cubic stiffness, respectively. For the linear case the resonance frequency is $\omega _{n}=6\textup{rad}/\textup{s}$. The linear case shows that the magnitude of the external excitation has no effect on the amplification gain, hence the resonance frequency of the system. But in the nonlinear case, the amplification gain is a function of both the magnitude and frequency of the input excitation, which causes a shifting in its resonance frequency for different values of $a$ and $\omega $. The active control scheme can adapt to these nonlinear effects. The closed-loop system is
\begin{equation}
m\ddot{q}+c\dot{q}+kq+k_{c}q^{3}=w-\theta _{p}q-\theta _{d}\dot{q}
\label{cubu}
\end{equation}

\begin{figure}
       \centering
        \begin{subfigure}[h] {0.35\textwidth}
                \includegraphics[width=\textwidth]{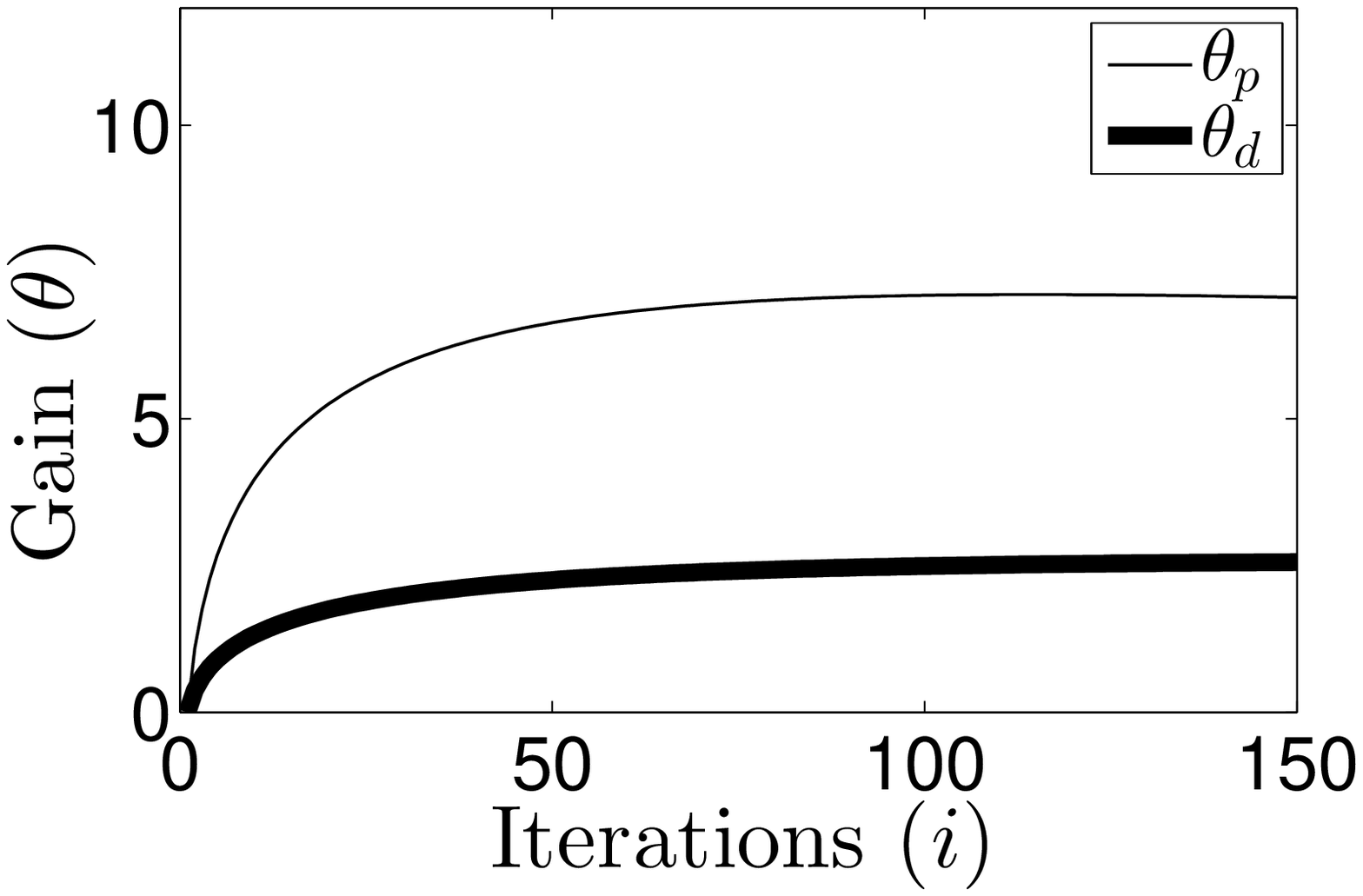}
                \caption{Evolution of adaptive gains}
                \label{gain}
        \end{subfigure}%
        \begin{subfigure}[h] {0.35\textwidth}
                \includegraphics[width=\textwidth]{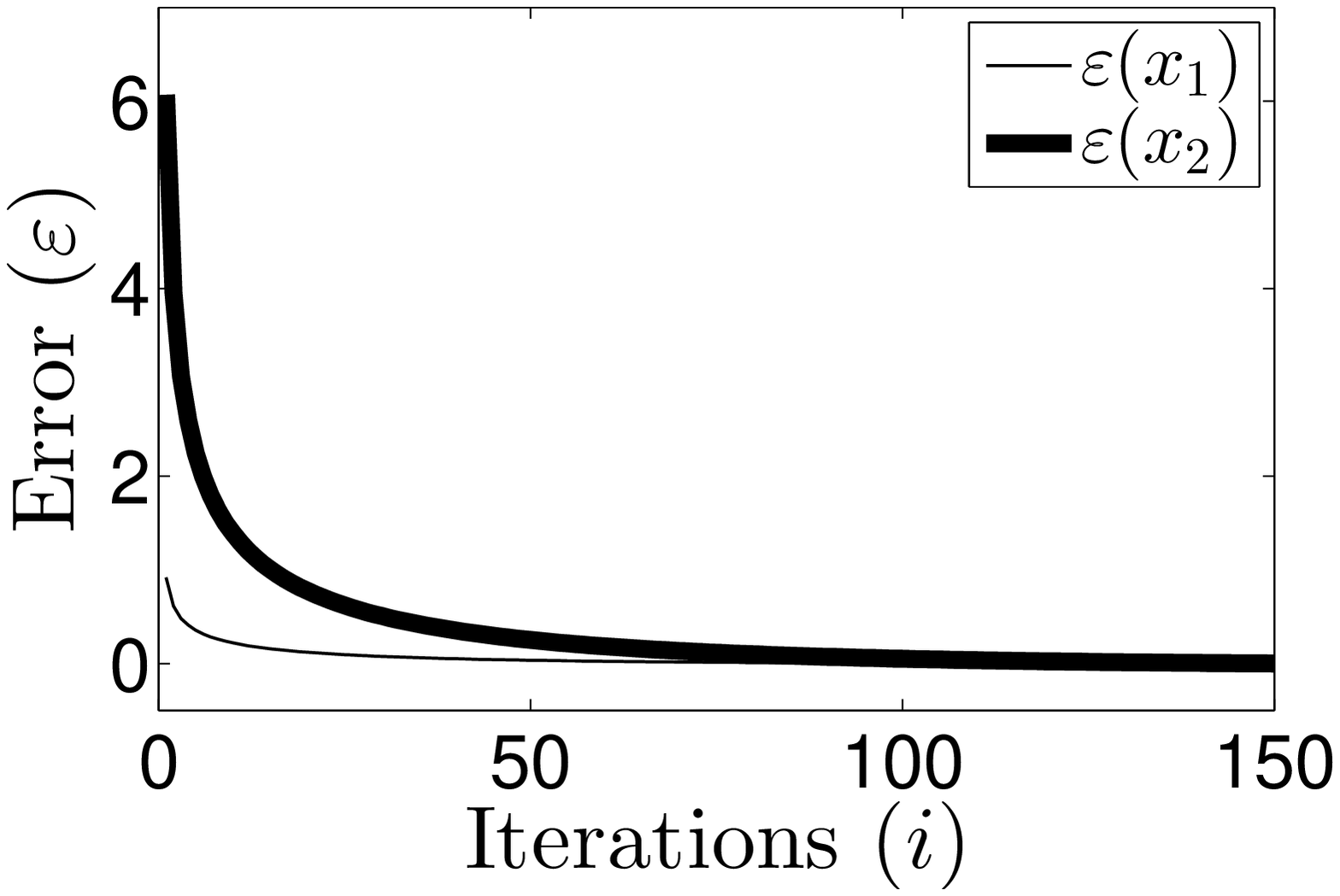}
                \caption{Evolution of adaptive error}
                \label{error}
        \end{subfigure}
       \caption{Evolution of adaptation parameters over each iteration.}\label{adapp}
\end{figure}

\begin{figure}[h]
\begin{center}
\includegraphics[width=7cm]{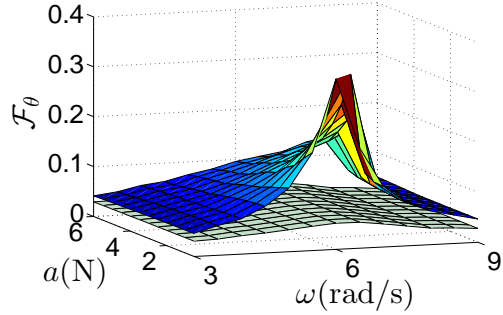}    
\caption{Uncontrolled and controlled $(\theta _{p}=7.1, \theta _{d}=2.6)$ FRF of the system.} 
\label{FRF}
\end{center}
\end{figure}

\begin{figure}
       \centering
        \begin{subfigure}[h] {0.4\textwidth}
                \includegraphics[width=\textwidth]{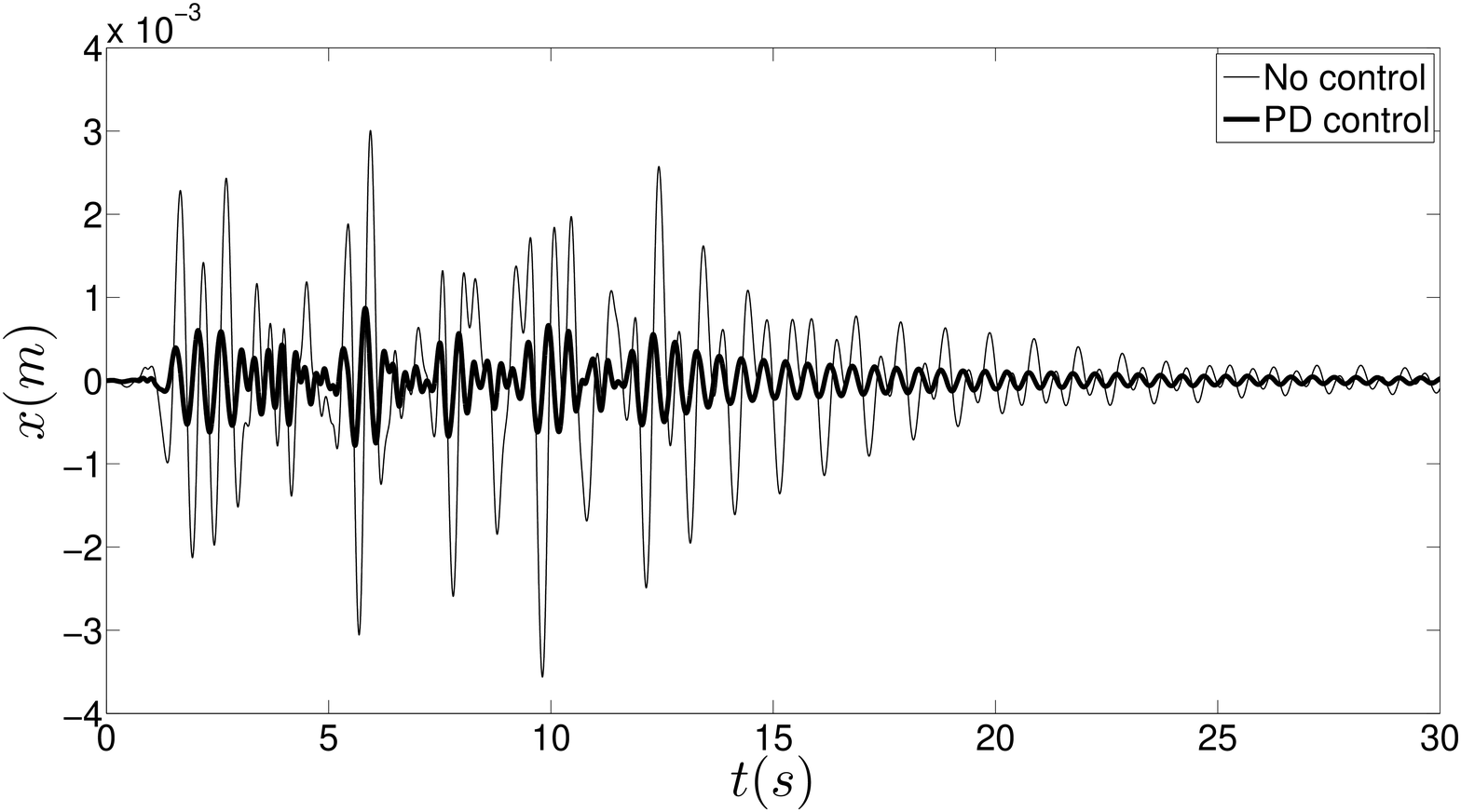}
                \caption{Uncontrolled and controlled displacements of the system.}
                \label{t_pd}
        \end{subfigure}%
        \begin{subfigure}[h] {0.4\textwidth}
                \includegraphics[width=\textwidth]{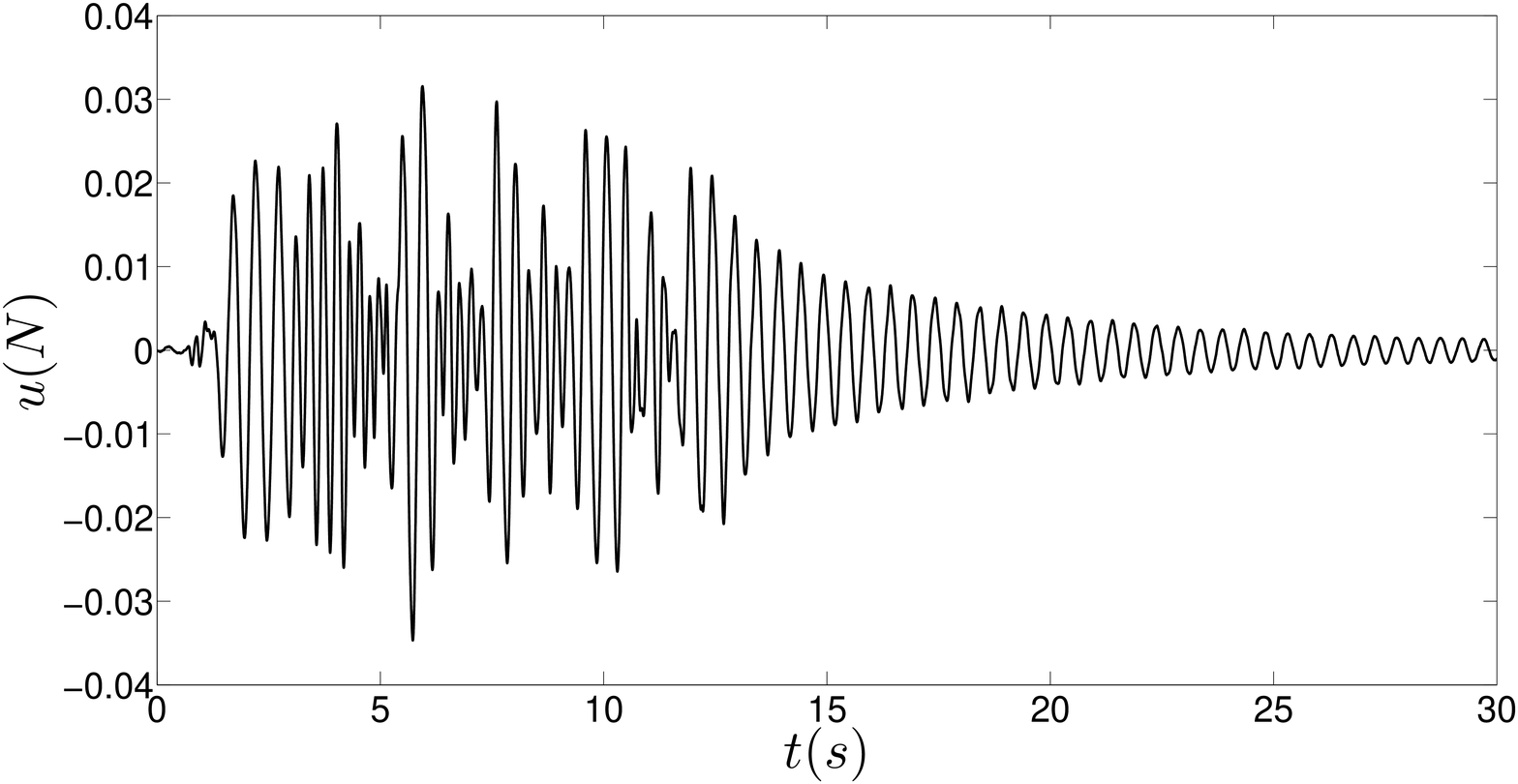}
                \caption{Control signal generated by the PD controller.}
                \label{u}
        \end{subfigure}
       \caption{Time response of the closed-loop system.}\label{time_r}
\end{figure}

The effects of adaptive PD controller on the performance of the active vibration control system are investigated. The adaptive algorithm parameters are set as: $\Gamma _{\theta _{p}}=120,\Gamma _{\theta _{d}}=1,\delta _{\overline{x}_{1}}=0.5,\delta _{\overline{x}_{2}}=3,$ and $\theta _{\min }=[0.001 \text{ } 0.001]$. The controller gains was started from a minimal value $\theta_{t_0} =\theta _{\min }$, and allowed to adapt using (\ref{adap}) after each set of FRF is calculated. Figure \ref{adapp}:(a) shows the controller gain adaptation process of the closed-loop system (\ref{cubu}). The corresponding error function $\varepsilon $ is shown in Figure \ref{adapp}:(b). Figure \ref{FRF} compares the FRF of the system without and with control, respectively. The F-norm of the system for position measurement, without ($\left\Vert \mathcal{F}_{0}(\overline{x}_{1})\right\Vert _{F}$) and with control ($\left\Vert \mathcal{F}_{\theta }(\overline{x}_{1})\right\Vert _{F}$) are $1.42$ and $0.5$, respectively. The controller response is obtained using the final values obtained from the adaptive scheme, i.e. $\theta _{p}=7.1\textup{N} /\textup{m},\theta _{d}=2.6 \textup{N} \textup{s}/\textup{m}$. Figure \ref{time_r}:(a) shows the time response of the displacements for both controlled and uncontrolled cases and the corresponding PD controller output signal is shown in Figure \ref{time_r}:(b).

From these simulation results, it is clear that the proposed algorithm assures an effective vibration attenuation over the band of excitation. Nonlinearities such as cubic stiffness can cause subharmonics in the system response. Since the F-norm of the system FRF is sensitive to any resonance in the excitation band, the iteration process is repeated until a stopping criterion is satisfied. Consequently, the new algorithm can effectively enhance the control performance over the given range of inputs. Furthermore, the stability analysis is much simpler due to the convergence analysis, which is obtained by performing simple coordinate transformation. Note that the controller updates its gains based on the measured input-output (amplification gain) relation of the system, which means that the exact knowledge of the system parameters is not required for the gain adaptation.

\subsection{Application to Satellite Attitude Control}

Momentum type actuators such as Reaction Wheels (RW) are widely used as attitude control actuators in spacecraft for orbital maneuvering. Unfortunately, these actuators are one of the main source of on-board vibration \cite{Masterson:2002}, which is caused by the static and dynamic imbalances in the RW assembly. This can be critical for the high precision space applications such as high-sensitivity imaging and astrometry. The dynamics and kinematics of a satellite system can be modeled as
\begin{equation}
H\dot{\omega}_{s}+S(\omega_{s}) H \omega_{s}=\tau
\label{sat}
\end{equation}
\begin{equation}
\dot{q}=J_s\omega_{s}
\label{kin}
\end{equation}
where $H \in \mathbb{R}^{3 \times 3}$ is the inertia matrix, $\omega_{s} \in \mathbb{R}^{3}$ is the angular velocity, $S(\omega_{s}) H$ is the angular momentum with $S(.)$, a skew-symmetric matrix representing the vector cross product, $\tau \in \mathbb{R}^{3}$ is the torque applied to the satellite system, $q \in \mathbb{R}^{3}$ is the satellite attitude vector, and $J_s \in \mathbb{R}^{3 \times 3}: \omega_{s} \rightarrow \dot{q}$ is the Jacobian matrix, all expressed in the satellite body frame. In this paper, the Modified Rodrigues parameters were used to represent the kinematic equations of the satellite system. 

By considering $q$ and $\dot{q}$ as the state-space coordinates and using (\ref{kin}), the equation of motion of the satellite system (\ref{sat}) can be written in the Lagrangian form as
\begin{equation}
{H}_{s}(q) \ddot{q}+ C_{s}(q, \dot{q}) \dot{q}=\tau_{s}
\label{nonc}
\end{equation}
where 
\begin{eqnarray*}
H_{s}(q) &=& J_{s}^{-T} H J_{s}^{-1}\\
C_{s}(q, \dot{q}) &=&J_{s}^{-T} H J_{s}^{-1} \dot{J}_{s} J_{s}^{-1} -J_{s}^{-T} S(\omega_{s}) H J_{s}^{-1} \\
\tau_{s}&=& J_{s}^{-T} \tau
\end{eqnarray*}

Since the system (\ref{nonc}) verifies the structure and skew-symmetric property \cite{Slotine:1990} of (\ref{nc}), it satisfies the theoretical analysis presented in the Section \ref{snc}. 

\begin{figure}[h]
\begin{center}
\includegraphics[width=10cm]{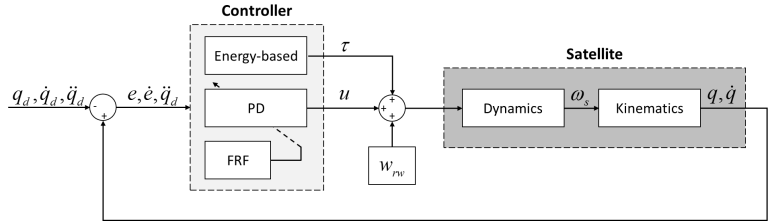}    
\caption{Block diagram of the active vibration control system for satellites.} 
\label{blocks}
\end{center}
\end{figure}

The rotational elements of the RW generates periodic disturbances, which can be modeled as a series of discrete harmonics \cite{Masterson:2002}
\begin{equation}
w_{rw}=\sum \limits_{i=1}^{\overline h} A_i\Omega^2 \text{ sin} \left(2\pi h_i \Omega t+\alpha _i\right)
\label{wrm}
\end{equation}
where $\overline h$ is the number of harmonics, $A_i$ is the amplitude coefficient of the $i$-th harmonic, $\Omega$ is the wheel speed, $h_i$ is the $i$-th harmonic number and $\alpha _i$ is a random phase. Since our disturbance model $w$ is similar to $w_{rw}$ by choosing $a=A_i\Omega^2$ and $\omega=2\pi h_i \Omega$, the proposed method can be a potential candidate in designing a vibration controller for the spacecraft systems with RW assembly. Now the satellite model with the disturbance $w_{rw}$ can be represented as
\begin{equation}
H\dot{\omega}_{s}+S(\omega_{s}) H \omega_{s}=\tau+u+w_{rw}
\label{satw}
\end{equation}
where $u$ is generated by the proposed FRF-based adaptive control algorithm for minimizing the vibration signals caused by the RW assembly. The block diagram of the implementation of the proposed algorithm is shown in Figure \ref{blocks}. 

Combining equations (\ref{ute}) and (\ref{nonc}) yields,
\begin{equation}
{H}_{s}(q) \ddot{q}+ C_{s}(q, \dot{q}) \dot{q}={H}(q) \ddot{q}^r+ C(q, \dot{q}) \dot{q}^r - (K_{r}+\Theta _r)(\dot{q}-\dot{q}^r)
\label{clconv}
\end{equation}
where $\ddot{q}^r=\ddot{q}^d-\Lambda _r \dot{e}$, and the corresponding closed-loop dynamics can be written in the following simple form 
\begin{equation}
{H}_{s}(q) \dot{r}+ \left( C_{s}(q, \dot{q})+K_{r}+\Theta _r\right)r=0
\label{clconv1}
\end{equation}

In order to implement the controller (\ref{ue}), we need to have the information regarding the parameters of the satellite. The satellite parameters utilized in \cite{Lizarralde:1996} are used here to illustrate the numerical results. The initial conditions are set as $q^0=\dot{q}^0=[0 \text{ } 0 \text{ } 0]^T$. The satellite system is excited using sinusoidal disturbances, and the corresponding position and velocity error $\left( \mathcal{F}_{\Theta}\left( \overline{e} \right) \text{ and } \mathcal{F}_{\Theta}\left( \dot {\overline{e}}\right) \right)$ at steady-state were used to obtain the FRF. The magnitudes and frequencies of the disturbance $w_{rw}$ for different RW velocities can be estimated as proposed in \cite{Masterson:2002}. During the tuning stage, the desired input $q_d$ is selected such that $q$ and $\dot{q}$ reaches steady-state. In this case, the satellite desired attitudes are chosen to be $q^d=[1 \text{ } 0.5 \text{ } 0]^T$. The adaptive algorithm parameters are set as: $\Gamma _{\Theta _{p}}=\text{diag}[10 \text{ } 10 \text{ } 10], \text{ } \Gamma _{\Theta _{d}}=\text{diag}[1000 \text{ } 1000 \text{ } 1000], \text{ } \delta _{\overline{e}}=[0.2 \text{ } 0.1 \text{ } 0.1], \text{ } \delta _{\dot {\overline{e}}}=[0.02 \text{ } 0.02 \text{ } 0.02],$ and $\theta _{\min }=0.001$. 

\begin{figure}
       \centering
        \begin{subfigure}[h] {0.33\textwidth}
                \includegraphics[width=\textwidth]{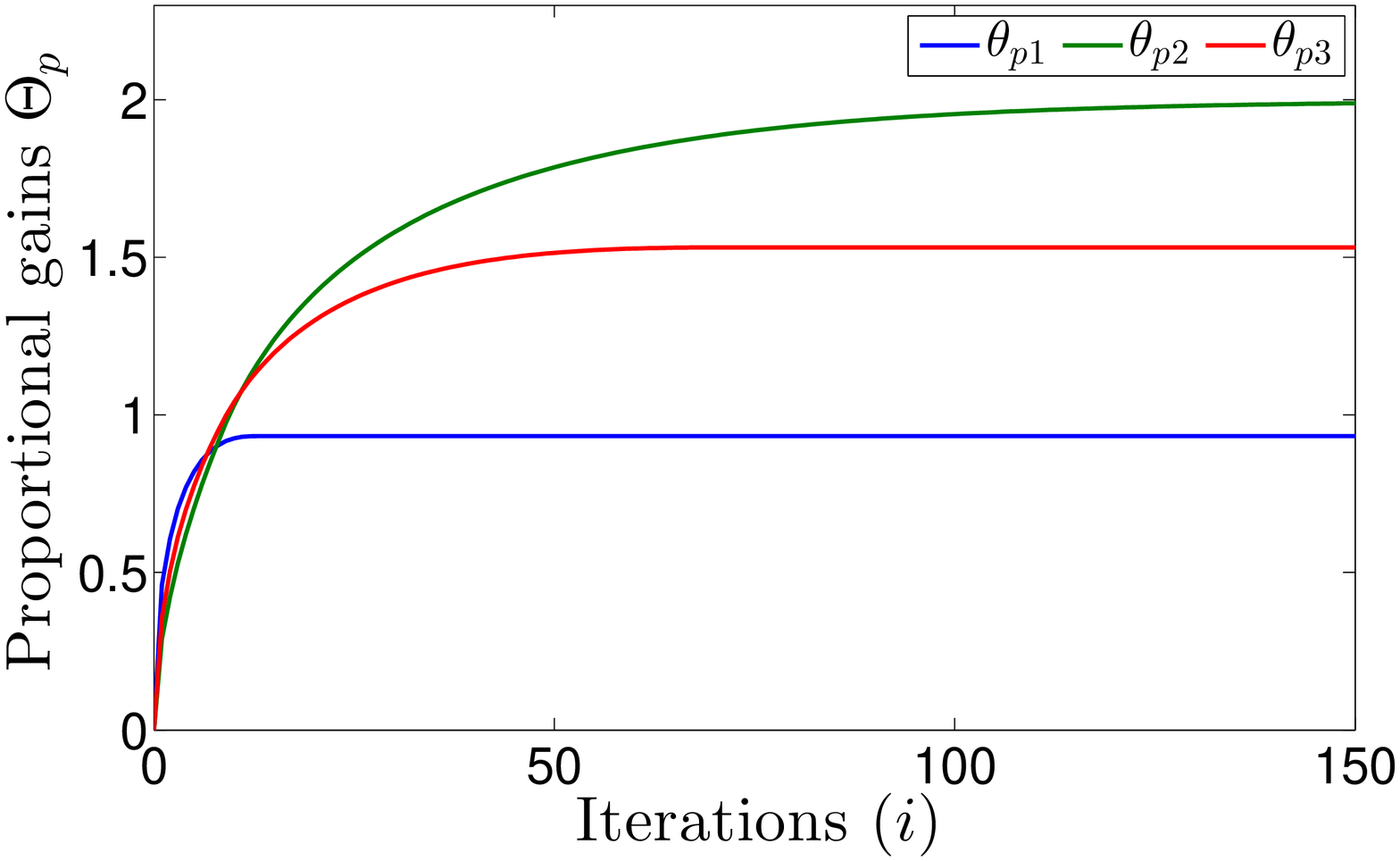}
                \caption{Proportional gain}
                \label{kpp}
        \end{subfigure}%
        \begin{subfigure}[h] {0.33\textwidth}
                \includegraphics[width=\textwidth]{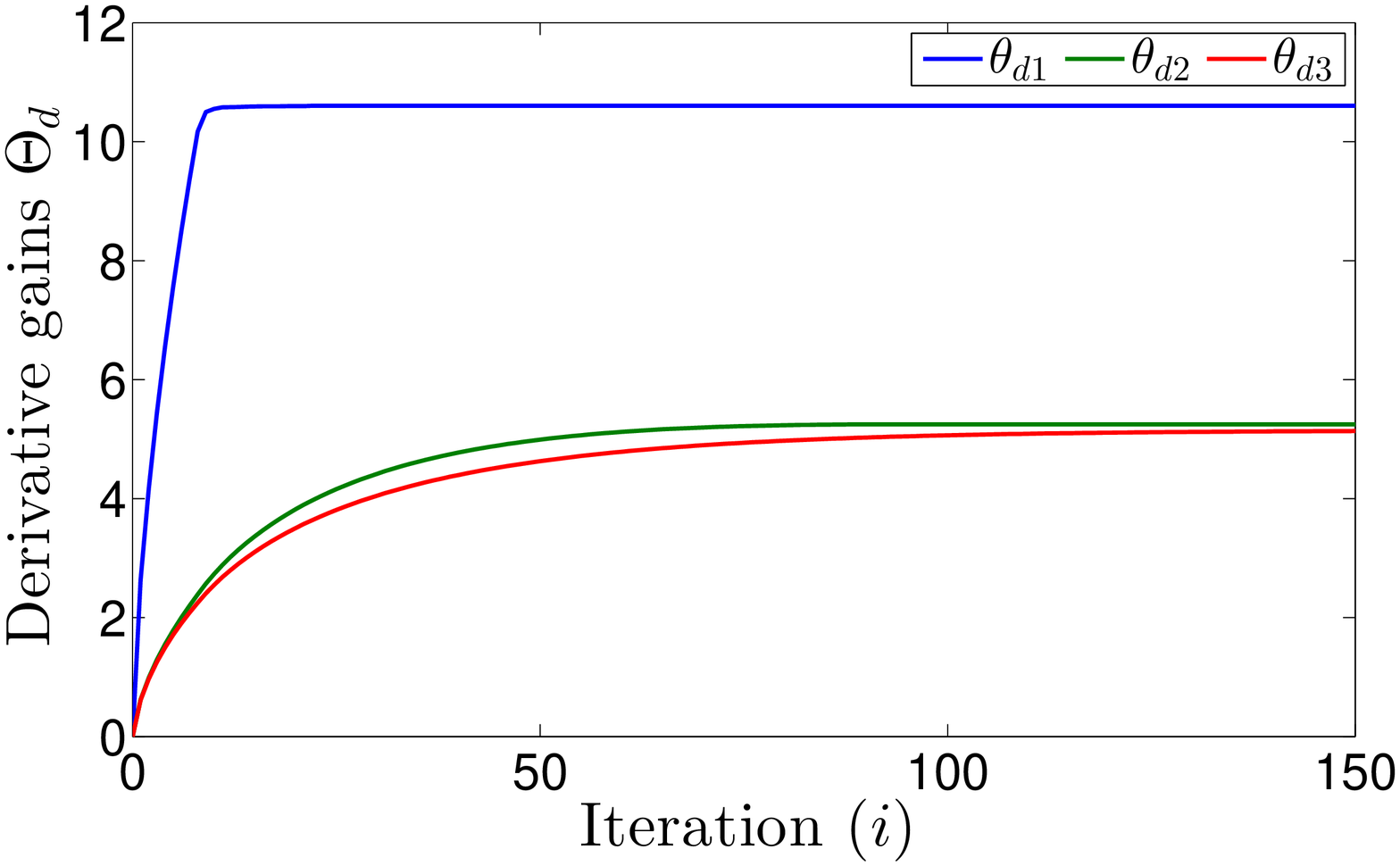}
                \caption{Derivative gain}
                \label{kdd}
        \end{subfigure}
       \caption{Evolution of adaptive gains.}\label{gains}
\end{figure}

\begin{figure}
       \centering
        \begin{subfigure}[h] {0.35\textwidth}
                \includegraphics[width=\textwidth]{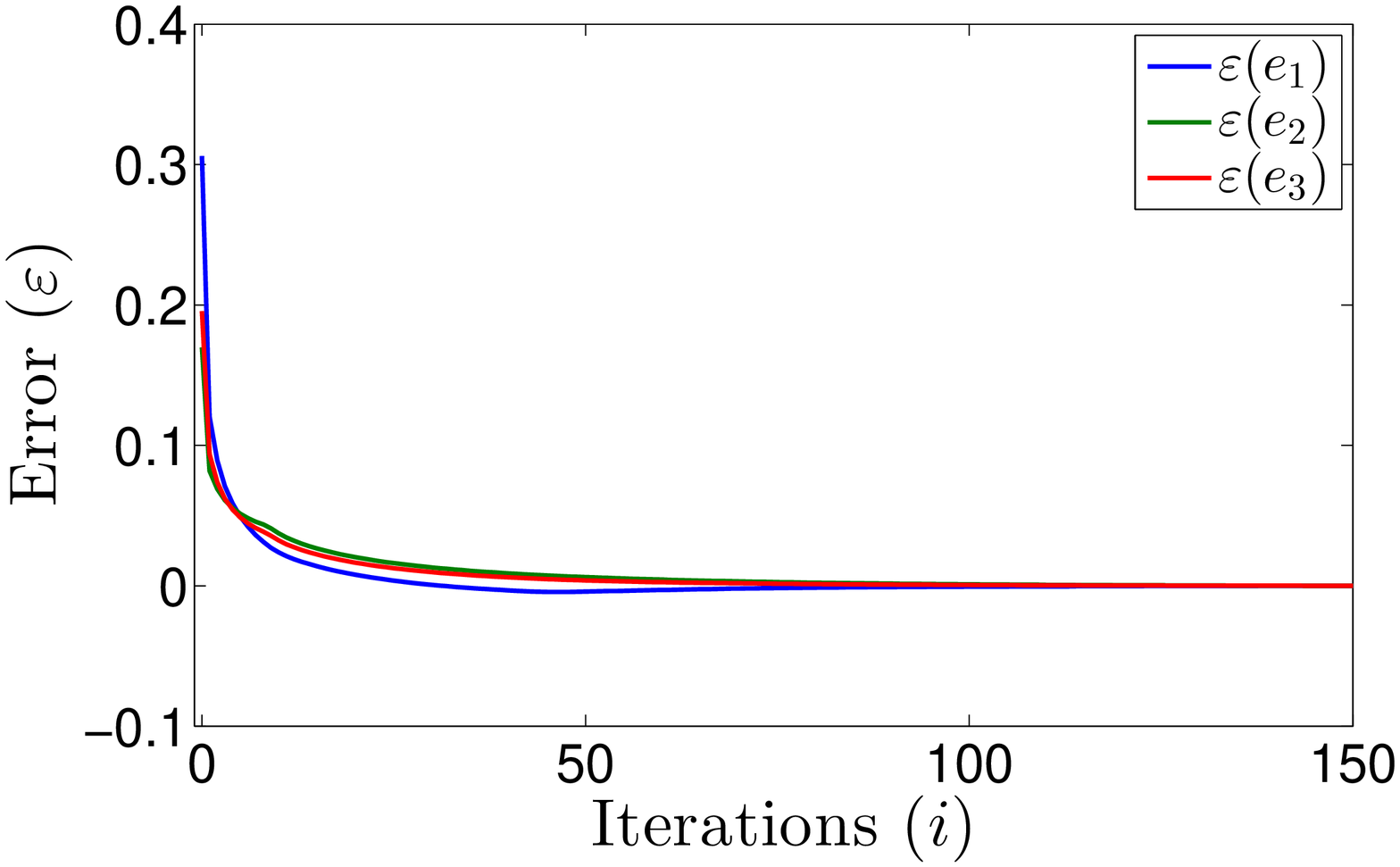}
                \caption{Proportional gain error}
                \label{ex}
        \end{subfigure}%
        \begin{subfigure}[h] {0.35\textwidth}
                \includegraphics[width=\textwidth]{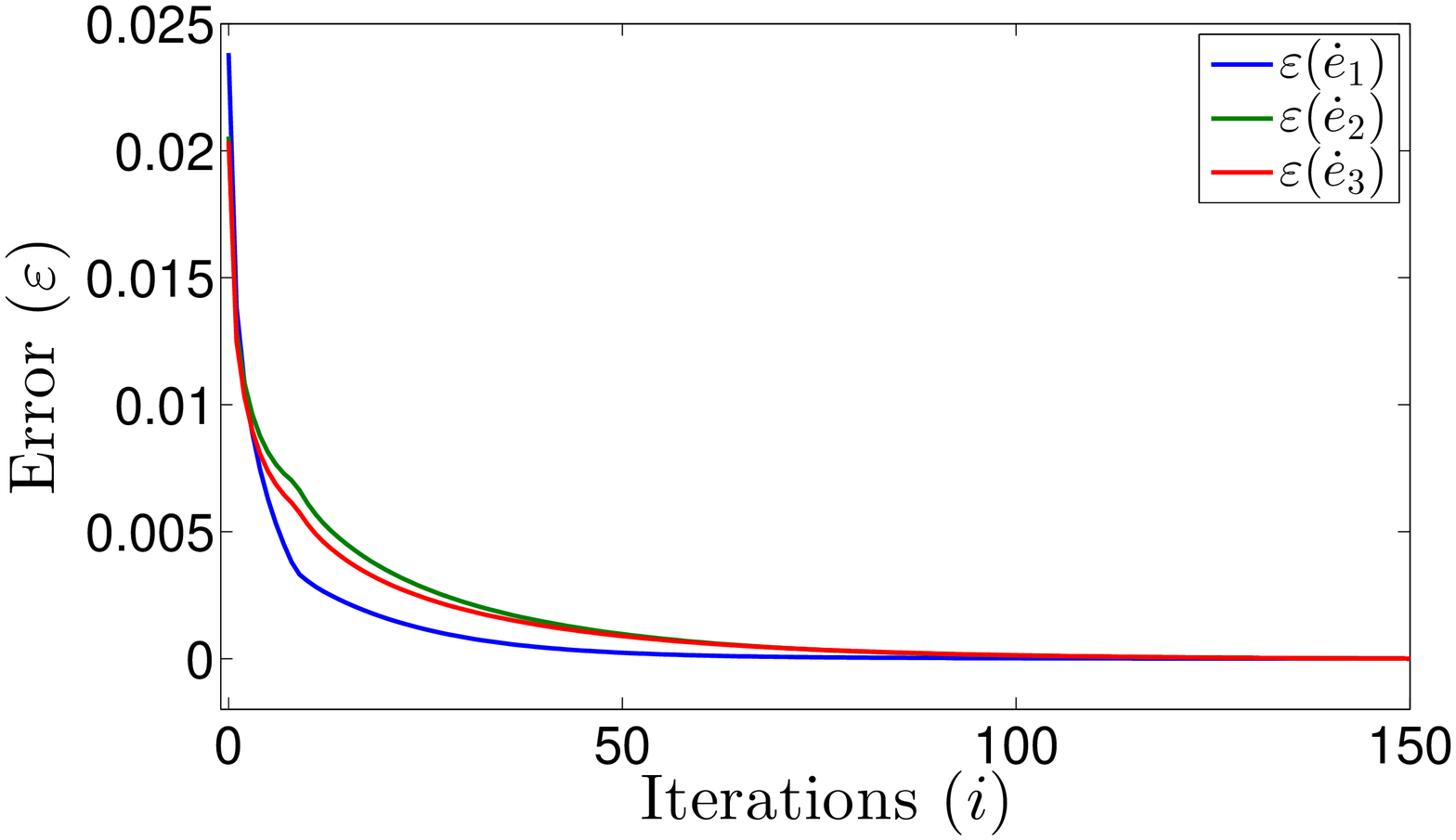}
                \caption{Derivative gain error}
                \label{ev}
        \end{subfigure}
       \caption{Evolution of adaptive errors.}\label{gain_err}
\end{figure}

\begin{figure}
       \centering
        \begin{subfigure}[h] {0.35\textwidth}
                \includegraphics[width=\textwidth]{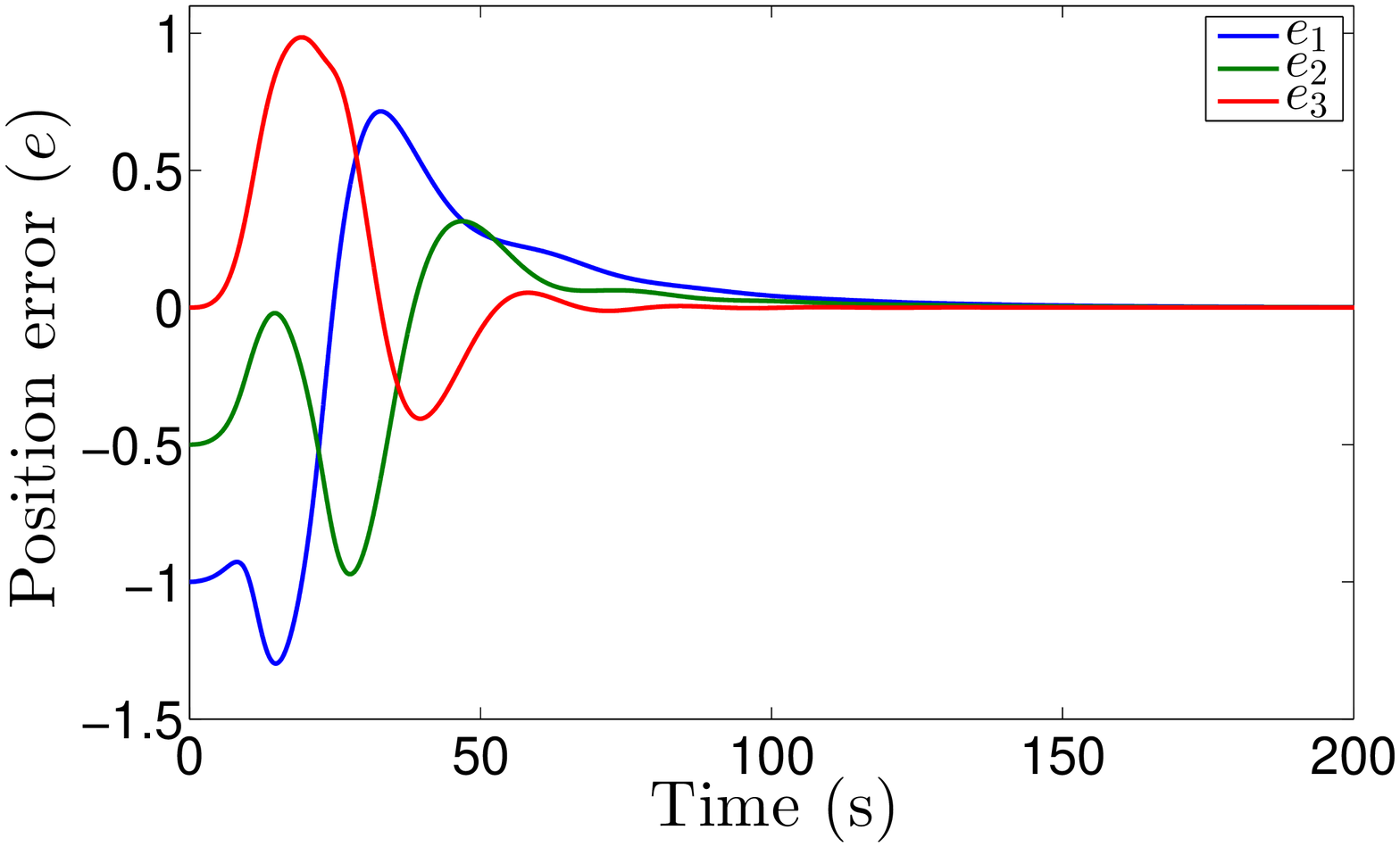}
                \caption{Position error}
                \label{e0}
        \end{subfigure}%
        \begin{subfigure}[h] {0.35\textwidth}
                \includegraphics[width=\textwidth]{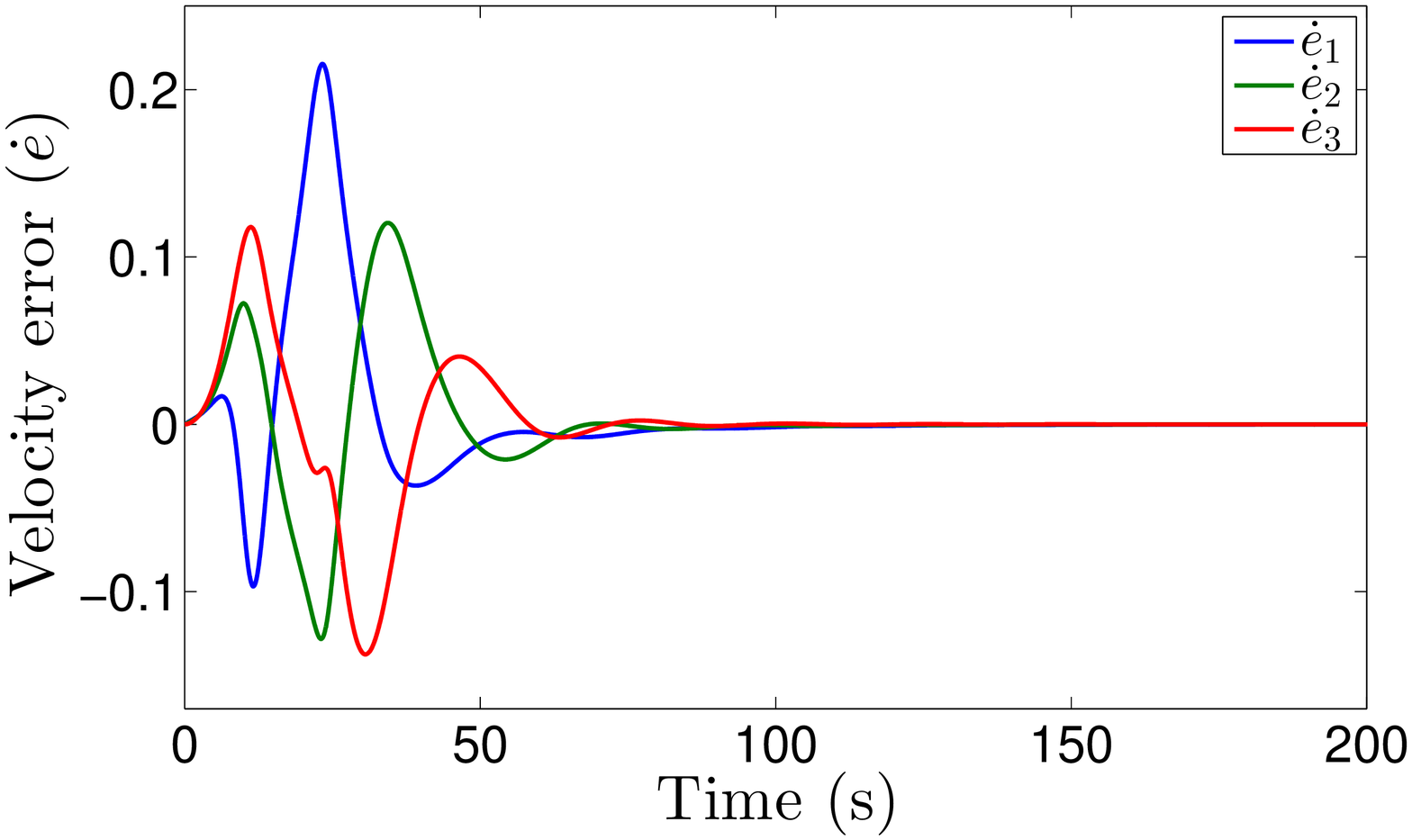}
                \caption{Velocity error}
                \label{de0}
        \end{subfigure}
       \caption{Without FRF based adaptive controller.}\label{err_0}
\end{figure}

\begin{figure}
       \centering
        \begin{subfigure}[h] {0.35\textwidth}
                \includegraphics[width=\textwidth]{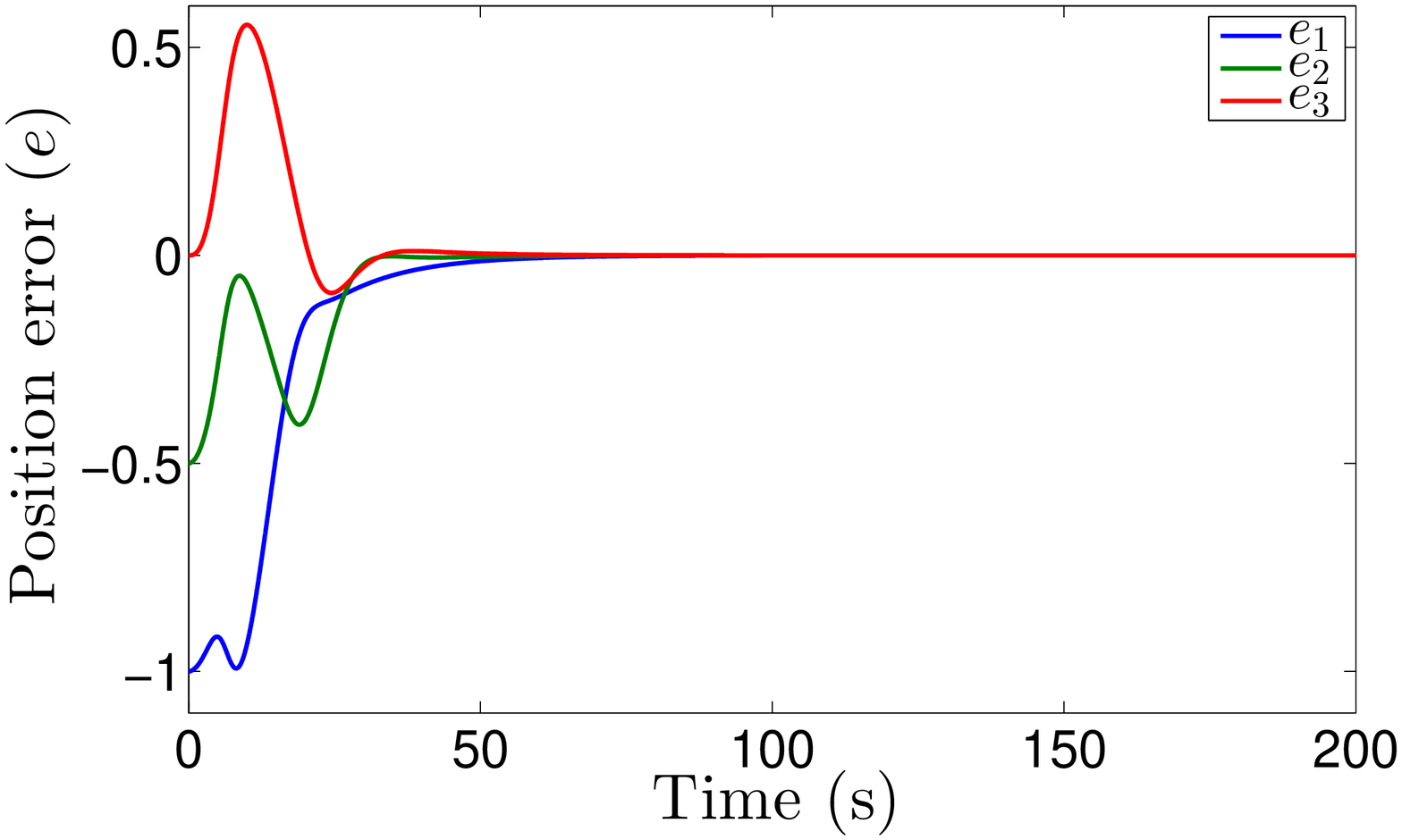}
                \caption{Position error}
                \label{eu}
        \end{subfigure}%
        \begin{subfigure}[h] {0.35\textwidth}
                \includegraphics[width=\textwidth]{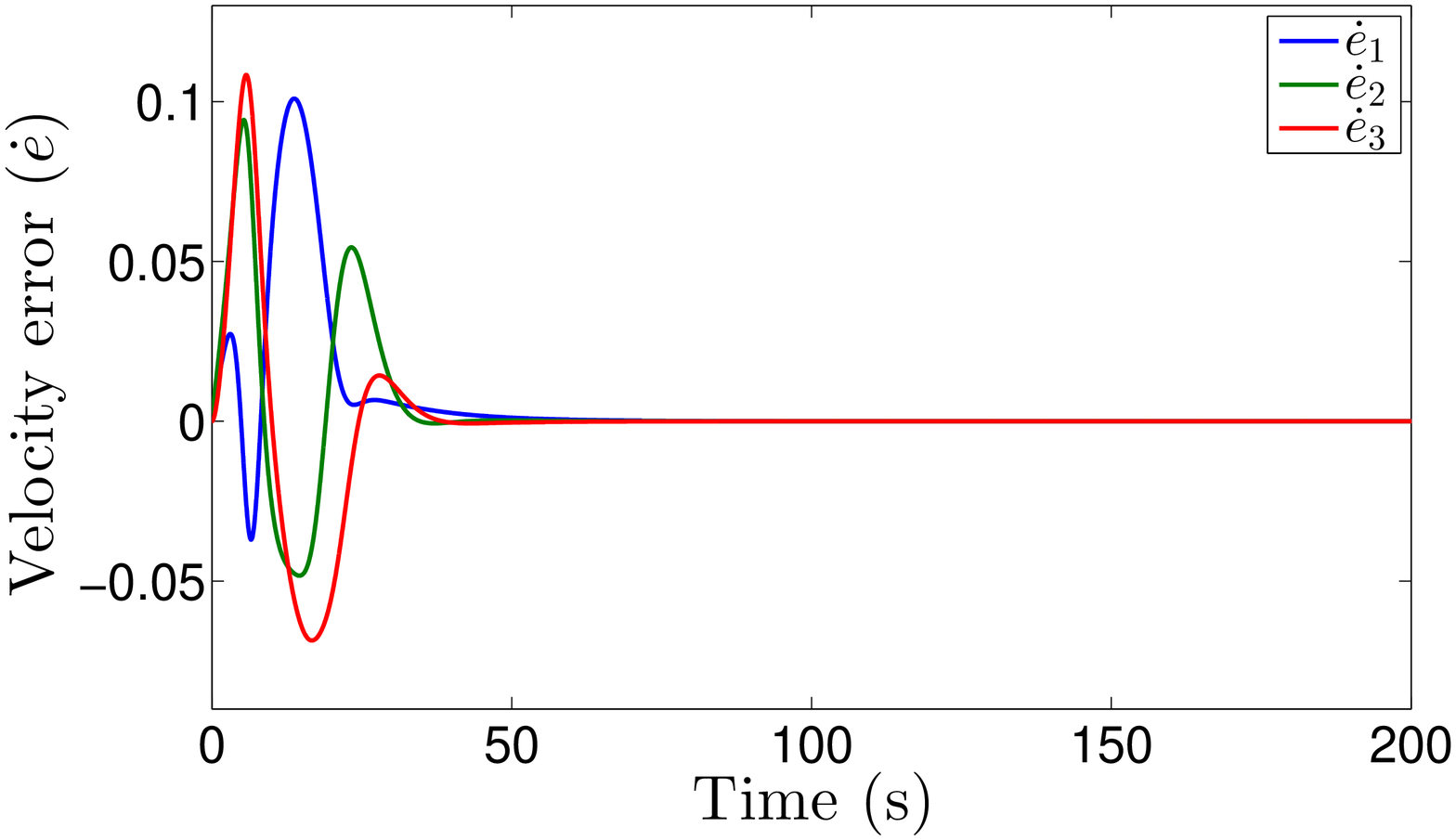}
                \caption{Velocity error}
                \label{deu}
        \end{subfigure}
       \caption{With FRF based adaptive controller.}\label{err_U}
\end{figure}

The adaptation of the proportional and derivative gains based on the FRF are shown in Figure \ref {gains} and the corresponding evolution of the errors, $\left(\left\Vert \mathcal{F}_{\Theta ,i}\left( \overline{e}\right)\right\Vert _{F}-\delta _{\overline{e}}\right)$ and $\left(\left\Vert \mathcal{F}_{\Theta ,i}\left( \dot{\overline{e}}\right)\right\Vert _{F}-\delta _{\dot{\overline{e}}}\right)$ are shown in Figure \ref {gain_err}:(a) and (b), respectively. It is seen that, by using the controller (\ref {ue}) the position and velocity error converges to zero quickly. However, the controller performance degrades in the presence of any excitation inputs. Figure \ref {err_0} shows the positional and velocity error of the satellite in the presence of excitation input, without using the adaptive controller ($u=0$). From these diagrams we can see that the controller requires some time interval to bring the error to zero. Figure \ref {err_U} shows the positional and velocity error of the satellite in the presence of excitation input, using the FRF based adaptive controller ($\Theta _{p}=\text{diag}[0.93 \text{ } 1.98 \text{ } 1.53], \text{and } \Theta _{d}=\text{diag}[10.60 \text{ } 5.24 \text{ } 5.13]$). From these diagrams, it is seen that the FRF based adaptive controller provides a suitable proportional and derivative gain such that the closed-loop system has less sensitivity towards the excitation force, hence induces the error caused by the excitation to converge quickly and as a result a better attitude control performance is achieved.

\section{Conclusions}

This paper has described an adaptive vibration control scheme for a class of nonlinear mechanical systems, subject to external excitation by using the nonlinear FRF technique. The nonlinear system possesses complex behavior for different excitation inputs. A simple and effective tool for vibration analysis and controller design has been proposed. Theoretically established that a mechanical system with an odd polynomial nonlinearity is convergent and hence a FRF can be derived. Furthermore, the stability is assured due to its convergence property. For the systems that are not convergent, a controller is used to establish the convergence. The main advantage of the frequency-domain approach presented in this article is that a satisfactory vibration attenuation for a band of excitation is assured. Finally, the proposed scheme is applied to the active vibration control problem of building structures with cubic nonlinearity and satellite systems. From the numerical studies, it is observed that the procedure is successful in suppressing the vibration. 





\section*{Acknowledgments}
\label{Ack}
This work is supported by PAPIIT of UNAM, Mexico under project IN113615. The first author would like to thank DGAPA of UNAM, Mexico for the Post-Doctoral fellowship.

\end{document}